\pdfoutput=1
\documentclass[pdflatex,sn-mathphys-num]{sn-jnl}

\usepackage{booktabs}

\usepackage{threeparttable}
\usepackage{float}
\usepackage[normalem]{ulem}
\usepackage[section]{placeins}
\usepackage{siunitx}
\usepackage{colortbl}
\usepackage{adjustbox}
\usepackage{makecell}

\usepackage{graphicx}
\usepackage{multirow}
\usepackage{amsmath,amssymb,amsfonts}
\usepackage{doi}
\usepackage{amsthm}
\usepackage{mathrsfs}
\usepackage[title]{appendix}
\usepackage{xcolor}
\usepackage{textcomp}
\usepackage{manyfoot}
\usepackage{algorithm}
\usepackage{algorithmicx}
\usepackage{algpseudocode}
\usepackage{listings}
\usepackage{enumitem}
\usepackage{url}

\newcommand{\pheno}{\textsc{PhenoLink}}
\newcommand{\phenseq}{\textsc{PhenoSeq}}
\newcommand{\DL}{\mathrm{DL}}

\newcommand{\best}[1]{\underline{\textbf{#1}}}

\begin{document}

\title[An Interaction Language Model]{An Interaction Language Model: Mechanism Discovery from Statistical Patterns of Physical Interactions}


\author[1]{\fnm{Triparna} \sur{Ganguly}\email{tliu@uga.edu}\email{yohannes.abate@uga.edu}}
\author[1]{\fnm{Hanqi} \sur{Jiang}}
\author[1]{\fnm{Xinliang} \sur{Li}}
\author[1]{\fnm{Yi} \sur{Pan}}
\author[1]{\fnm{Junhao} \sur{Chen}}
\author[2]{\fnm{Miyuki} \sur{Karunathilaka}}
\author*[1]{\fnm{Tianming} \sur{Liu}}
\author*[2]{\fnm{Yohannes} \sur{Abate}}

\affil[1]{\orgdiv{School Of Computing}, \orgname{University of Georgia}, \orgaddress{\city{Athens}, \state{GA}, \postcode{30602}, \country{USA}}}

\affil[2]{\orgdiv{Department of Physics \& Astronomy}, \orgname{University of Georgia}, \orgaddress{\city{Athens}, \state{GA}, \postcode{30602}, \country{USA}}}
\abstract{
Interactions among elementary building blocks across physical, chemical, and biological systems follow structured patterns that can be interpreted as a learnable language: just as natural language models learn which words tend to follow others, one can learn which physical phenomena tend to follow others and under what conditions. We introduce an Interaction Language Model (ILM): a framework that treats interaction as a statistically structured, learnable language for design-to-function reasoning across scientific fields, with a deliberate architectural commitment to keeping statistical chain proposal separate from corpus-grounded chain verification. By capturing statistical dependencies and ordered relationships among interacting components and verifying every proposed transition against a fixed evidence graph, ILMs can infer interaction pathways, identify missing steps, and propose new functions while exposing a per-step audit trail that is reproducible from the same artifact. We demonstrate this framework through two complementary components, which we call PhenoLink and PhenoSeq, applied here to molecular diffusion, a ubiquitous mechanism that governs energy and particle transport across natural and engineered systems. PhenoLink is a directed graph of interactions among diffusion based events extracted from scientific publications, in which each edge aggregates paragraph-level evidence and carries a transition probability interpreted as an information cost. PhenoSeq complements this structure with a sequence model that proposes missing mechanistic steps under endpoint constraints, yielding a generate-then-verify pipeline in which every accepted step carries a reproducible, corpus-grounded audit trail. 
Together they form a generate-then-verify
pipeline that returns every explanation with a reproducible, per-step audit trail.
Across four held-out test suites of 172 queries, the PhenoLink–PhenoSeq
pipeline matches a zero-shot Claude Opus 4.7 baseline on in-distribution accu-
racy, yet unlike the baseline, which fabricates a chain for every impossible
query, it refuses up to 93.5\% of them and grounds 100\% of its accepted tran-
sitions in corpus evidence by construction. ILM thus delivers what zero-shot
generation structurally cannot: deterministic refusal of evidence-absent queries
and auditable, per-step support. The framework can be extended to a wide range of natural and engineered systems where function emerges from sequential and conditional interactions among molecules, cells, devices, or other fundamental components.
}

\maketitle

%

%

\section{Introduction}\label{sec:intro}
Interactions among elementary components can give rise to functional units in natural and engineered systems. These interactions follow structured patterns governed by physical laws, where certain transitions between states are warranted under specific conditions and others are not. We observe that these patterns share a fundamental property with natural language: they are sequential, context-dependent, and statistically learnable. Just as language models learn which words tend to follow others, one can learn which physical phenomena tend to follow others, and under what conditions. Motivated by this observation, we introduce the Interaction Language Model (ILM): a framework that captures these statistical dependencies between interacting phenomena and uses them for prediction of interaction pathways, identification of missing steps, and auditing of unsupported transitions. The architectural feature of the ILM framework is a deliberate separation between statistical \emph{proposal} and corpus-grounded \emph{verification}: candidate reasoning chains are generated by a sequence model whose outputs are independently checked, step by step, against a precompiled directed graph of evidence-weighted phenomenon transitions, so that every accepted explanation is reproducible from a fixed graph artifact and queries lacking corpus support are refused rather than completed.

We note that LLMs can generate domain-specific reasoning for interaction towards a function but may embed unsupported transitions that are difficult to detect without domain expertise~\cite{ji2023hallucination,huang2025llmsurvey}. Domain fine-tuning improves recall but does not provide per-step evidence mapping.
The ILM framework we propose addresses this shortcoming by starting from a fundamentally different perspective: we separate generation from support, constructing reusable statistical knowledge objects that can audit any candidate explanation. The framework consists of two complementary components we call \pheno{} and \phenseq{}. \pheno{} is a directed graph of phenomenon-to-phenomenon interactions extracted from scientific publications. We use the term \emph{interaction} here in a specific, operational sense: a directed mechanistic transition in which one phenomenon causally or conditionally precedes another under specified physical conditions, with the precedence supported by paragraph-level corpus evidence. This usage is distinct from the symmetric, entity-level interactions familiar from systems biology (e.g., protein--protein binding, gene--gene co-expression, or regulator--target associations), which typically connect named entities rather than physical processes and do not encode causal direction. In \pheno{}, each edge aggregates paragraph-level evidence and carries a transition probability interpreted as an information cost~\cite{rissanen1978mdl,grunwald2007mdl}. \phenseq{} complements this structure with a sequence model~\cite{sutskever2014seq2seq,vaswani2017attention} that proposes missing mechanistic steps under endpoint constraints. Together, they support a pipeline in which candidate explanations are generated, checked step by step against corpus-grounded evidence, and returned with transparent audit trails.

The key distinction between ILM and standard LLMs lies not in raw pass rate on well-posed scientific queries, where modern LLMs are competitive on domains such as molecular diffusion that are well-represented in training corpora~\cite{li2024scibench,sun2024scieval}, but in two structural properties of the resulting reasoning. First, a standard LLM has no internal representation of which transitions are supported by corpus evidence and which are not. Asked an impossible question, it will produce a fluent chain anyway: in our evaluation, vanilla Claude Opus 4.7 fabricated a mechanistic chain for every one of 31 deliberately-constructed missing-edge queries, never refusing once. Standard LLMs treat phenomenon names as text tokens; ``temperature gradient,'' ``diffusion,'' and ``mass transfer'' are sequences of subword units whose relationships are learned implicitly from co-occurrence in training corpora and have no concept of evidence absence to refuse on. Second, even when a standard LLM produces a chain that is fluent and reasonably cited, the steps in that chain do not correspond to entries in any specific corpus or vocabulary; we observed that only 0.70\% of consecutive (u,v) transitions in vanilla Claude's chains match edges in our precompiled evidence graph, compared to 100\% by construction in \pheno{} chains. In contrast, the ILM framework tokenises the phenomena themselves, not the words that describe them, and learns evidence-weighted transition scores grounded in paragraph-level corpus evidence. Each accepted transition in the resulting graph carries an explicit evidence trail, with the supporting paragraphs identified at construction time and reproducible from the released artifact. We therefore frame the operational distinction between ILM and standard LLMs as (i) deterministic refusal of queries for which the evidence graph contains no supported path, and (ii) per-step audit values reproducible from a fixed artifact under Tiers~0--2 (Sec.~\ref{sec:methods_pipeline}), rather than as a claim of pass-rate superiority on standard queries.

 Several recent systems address LLM auditability gaps by augmenting LLMs with structured knowledge or retrieval at inference time. GraphRAG~\cite{c1799bf28d1ae93e1631be5b59196ee1e568f538} extracts an entity-level knowledge graph from a text corpus and uses community summaries to support global question answering; CausalRAG~\cite{71bb53ceb14affbac0811516dfa6d22f49f7dda6} integrates causal graphs into the retrieval pipeline; and a broader class of knowledge-graph-augmented LLM systems~\cite{pan2024kgllm} enables source-traceable answers. These systems are an important part of the design space, and our framework is best understood as a complementary point in it. The distinctive features of the Interaction Language Model (ILM) approach are: (i) a directed graph defined at the level of canonical \emph{phenomena} rather than free-text entities, in which every edge encodes an aggregated, evidence-weighted directional score; (ii) graph construction precompiled from the literature once, so per-step audit values are deterministic and reproducible across runs (in contrast to retrieval-augmented systems whose retrieved passages can vary across runs); and (iii) an explicit \emph{proposal/verification separation}, in which a sequence model proposes candidate chains and the graph independently verifies each step against corpus evidence under a fixed, transparent scoring rule. We evaluate this pipeline on 172 endpoint queries across four held-out test suites and compare it directly to a frontier large language model prompted zero-shot on the same queries; the comparison is summarised in Sec.~\ref{sec:results_external} and informs the framing of our contribution as architectural rather than benchmark-superiority.

To demonstrate the ILM framework, we focus on one especially ubiquitous class of interactions in science: those governed by particle diffusion. Molecular diffusion provides a rigorous test case because correct reasoning in this domain requires structured mechanistic knowledge~\cite{mason1983dusty,cussler2009diffusion,bird2002transport}, including which transport regime applies, which assumptions are valid, and which physical relationships justify each inferential step. Here, accuracy depends not only on knowing individual facts, but on determining whether transitions between concepts are physically warranted under the stated conditions and whether the next inferential step follows from the governing physics. Reasoning about diffusion is therefore inherently sequential, conditional, and mechanistic, making it an ideal model system for showing how ILM can move beyond static knowledge retrieval toward stepwise, physically grounded reasoning about interactions and function.
Here we make three specific contributions. First, we formalise phenomenon-level chain reasoning as a proposal/verification problem, separating a closed-vocabulary sequence proposer from a fixed evidence graph that can audit any proposed transition. Second, we construct and release an interaction graph (using molecular-diffusion interaction as a model example) with canonical phenomenon identifiers, directed edges, graph-normalised evidence scores, and paragraph-level support records. Third, we evaluate the framework not only on graph-consistent endpoint completion but also on refusal behaviour, out-of-vocabulary handling, missing-edge queries, and a zero-shot LLM comparison.
\section{Related work}\label{sec:methods_related}

\pheno{} occupies a deliberately narrow position in this landscape, between traditional knowledge graphs, retrieval-augmented generation, and neural sequence models. Like a knowledge graph, it stores entities and directed relations; unlike many knowledge graphs, each edge is associated with paragraph-level evidence, aggregated confidence, and a reusable audit record. Like retrieval-augmented systems, it grounds reasoning in scientific literature; unlike them, the grounding is precompiled into a structured graph rather than retrieved ad-hoc at inference time, so the per-step audit values returned for a given query are deterministic and reproducible across runs. Like a sequence model, \phenseq{} proposes ordered chains; unlike natural-language generation, the output vocabulary is restricted to canonical phenomena and every proposed transition is checked against graph evidence. \pheno{} therefore targets a restricted, phenomena-level state space and treats edges as auditable corpus-supported directed associations with per-step evidence storage, graph-normalised evidence scores, and chain-level path-cost scoring: features designed for reasoning-step inspection rather than global concept coverage.
 
Scientific knowledge graphs encode entities and relations from text and databases~\cite{hogan2021knowledge,pan2024kgllm}. Prior work on causal knowledge graphs from text has focused on recovering corpus-level causal structure over variables or events across diverse domains: scientific literature~\cite{0504c5d96ebf7be01e2b622ff3a4bf155f2b0a41,e9273e5c14343c0b34010bbf2e2aada57d5f6bcf}, semi-structured industrial records~\cite{9f262b2d065e4e0cfebfa8f2a56962278af65982}, financial text~\cite{5e39509f50e4c57f7906dad65c39b422924b26d0,97d66ce88e94dd3d15a6455fe22192d8e9f1c1a5,49d87269fbc2b1534dd2f60c9d9c669e4729d318}, news and web sources~\cite{35ca944aed7d01ef146c4b65b3e9dd7fb0233837,503dcb22246bd615f50e280b68339b8bd656a2a1,2911c478e8f97855486e0104bcc440455af3c40f,b8066ba19708adafc72bdb9f246c071ae400017a,c18f3023da76cdcb00939b7b2219ecbe4959a446}, and general causal reasoning~\cite{kiciman2023causal}. These works generally encode generic cause--effect edges between concepts and aim at broad coverage or decision support. Recent work has further explored LLMs for causal graph discovery, finding them effective as imperfect prior sources but prone to associative rather than mechanistic and evidence-grounded reasoning~\cite{kiciman2023causal}. \pheno{} differs in targeting a restricted, phenomena-level state space and treating edges as auditable corpus-supported directed associations with per-step evidence storage, graph-normalised evidence scores derived from aggregated corpus-paragraph extractions, and chain-level path-cost scoring: features designed for reasoning-step inspection rather than global concept coverage.
 
A rapidly expanding line of recent work couples large language models with explicit graph structure to make multi-step reasoning more reliable and inspectable, and it is against this backdrop that the contribution of the ILM framework is sharpest. Graph-augmented reasoning methods inject a retrieved or constructed graph into the model's reasoning process to improve complex inference~\cite{98411a0a6eed28ee6f5c2ff44cc7d0304e440243}; graph-augmented retrieval pipelines specialised for causal annotation~\cite{4cee4ef8aef17af2ed1bfe28fbb45e044482207f} or deductive fact verification~\cite{f7367694d236da62a91018815c3066ee20dc66d0} use causal knowledge graphs to supply traceable supporting context at inference time; and a complementary strand builds dedicated causal-chain reasoners, including event-graph-enhanced explainable reasoning~\cite{cab8ddc42cefbedc7ef8579888c2475e9609a5d7} and structural recurrent models for reliable causal-chain inference~\cite{0866bf053f0926e115dc2329d4da226a74b10e3e}, while still other work asks whether LLMs can validate proposed causal edges at all~\cite{0e48e746c316cff43d9e27d3061408ea6eee2aef}. These systems share our aim of grounding generated reasoning in explicit relational structure, yet they differ from ILM in two architecturally decisive ways. First, in graph- and retrieval-augmented LLMs the graph is consulted \emph{per query at inference time} to condition a generative model, so the retrieved context, and therefore the resulting explanation, can vary across runs; \pheno{} is instead compiled once from the corpus, so the per-step evidence score returned for a given transition is fixed and reproducible. Second, these systems typically interleave proposal and grounding within a single generative pass, whereas ILM enforces an explicit \emph{proposal/verification separation}: \phenseq{} proposes candidate chains over a closed phenomenon vocabulary and \pheno{} independently audits every transition against paragraph-level evidence under a transparent scoring rule, so a fluent-but-unsupported proposal can never be silently accepted. It is this separation, rather than raw generative capacity, that lets the framework refuse physically implausible chains and attach a deterministic audit trail to each accepted step.

\section{Methods}\label{sec:methods}

\subsection{Corpus and phenomenon vocabulary}\label{sec:methods_corpus}

To develop a comprehensive library of phenomena relevant to systems governed by molecular diffusion, we first assembled a large scientific corpus from published articles in high-impact journals. The full corpus consists of 983 PDF documents (${\sim}3.2$~GB), covering the period from 1903 to 2024 after filtering out spurious years inferred from filenames. Of these, 145 papers ultimately contributed at least one extracted edge to the released graph; the remaining papers either yielded no edges that survived the orientation and confidence filters, or were not processed in the released cache for cost reasons. We make this distinction explicit because reviewers can inspect the released cache to verify which papers contributed which edges. From this literature corpus, we constructed a controlled phenomena vocabulary, $\mathcal{V}$, containing 782 terms that capture concepts central to molecular diffusion and related transport behaviour. These terms span fundamental mechanisms, transport regimes, material descriptors, and governing physical quantities. Representative examples include mean free path, Knudsen diffusion, Fickian diffusion, effective diffusivity, tortuosity, porosity,
binary diffusion, Stefan--Maxwell diffusion~\cite{krishna1997maxwell}, transition regime, adsorption, and surface diffusion. A complete list of the vocabulary used in this work is provided in the Supporting Information (SI). To ensure consistency and machine readability, each phenomenon in the vocabulary is assigned a stable token identifier composed of letters and numbers (for example, A00017). The released artifact \texttt{vocab.json} provides a bijective mapping between phenomenon names and token identifiers, enabling reproducible indexing, retrieval, and downstream analysis.\\

\begin{figure*}[htbp]
    \centering
    \begin{adjustbox}{width=1\textwidth, center}
        \includegraphics{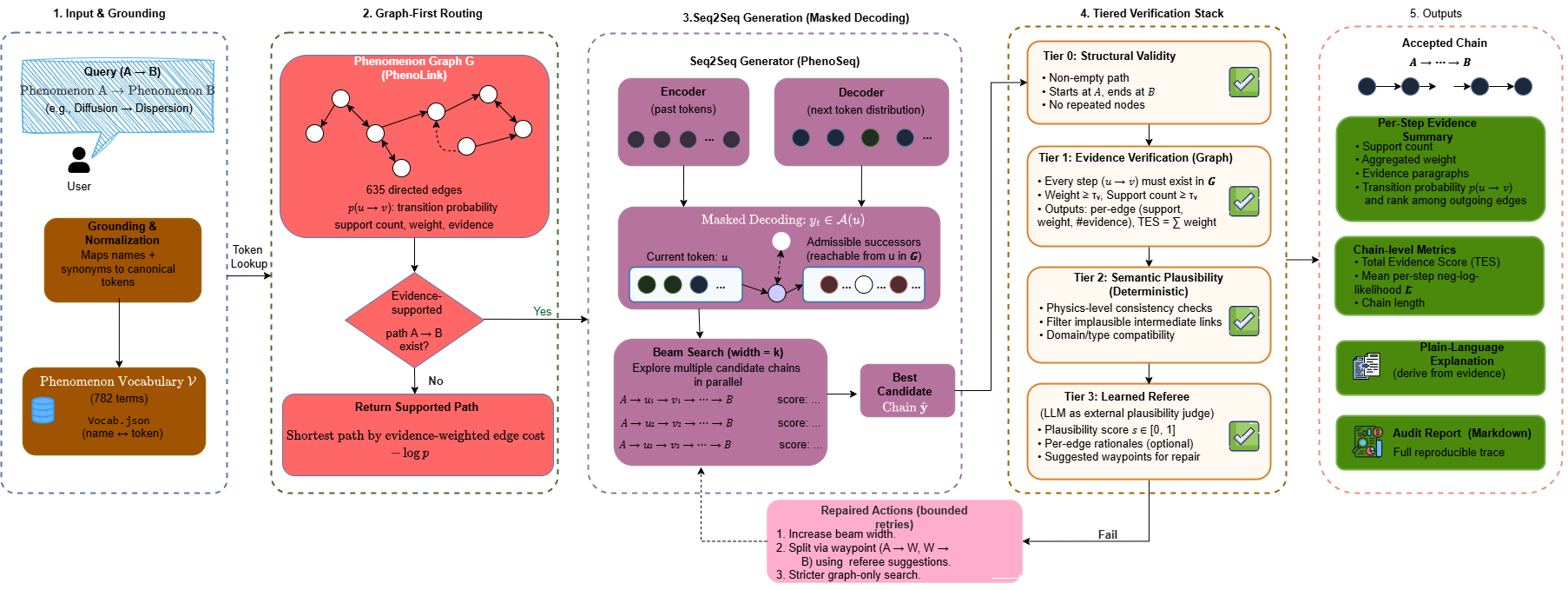}
    \end{adjustbox}
    \caption{Overview of the \pheno{}--\phenseq{} agentic reasoning pipeline. User queries specifying endpoint phenomena are grounded to canonical tokens through normalization and vocabulary lookup. If a supported graph path already exists, it is returned directly. Otherwise, the Seq2Seq generator (with masked decoding) proposes a candidate chain $A \rightarrow \cdots \rightarrow B$, which is then evaluated by a multi-tier verification stack. Tier~0 checks structure and endpoint consistency; Tier~1 verifies evidence support against \pheno{}; Tier~2 applies semantic plausibility filtering. Failed candidates trigger bounded repair actions (beam expansion, waypoint splitting, graph fallback). Tier~3 provides an optional learned referee for additional plausibility scoring and repair guidance. Final outputs consist of an accepted chain together with per-step evidence, transition probabilities, and a plain-language interpretation. The bottom panels detail the graph scoring mechanism (left), the tier verification checks (centre), and the OOV/referee handling pathways (right).}
    \label{fig:pipeline_overview}
\end{figure*}

Figure~\ref{fig:pipeline_overview} shows the complete ILM agentic reasoning pipeline, comprising five logical stages: input grounding, graph-first routing, seq2seq generation, tiered verification, and output. Given two endpoint phenomena, the system either returns a chain via direct graph routing (if a supported path exists) or via masked-decoding seq2seq proposal followed by a tiered verification stack. This architecture operationalises the central design commitment of the framework: statistical chain proposal is kept separate from corpus-grounded chain verification, so that every accepted explanation is reproducible from a fixed graph artifact. Failed candidates trigger bounded repair actions (beam expansion, waypoint splitting, graph-only fallback); the optional Tier-3 referee provides external plausibility scoring but lies outside the deterministic-audit core. The pipeline collectively produces, for every query, an accepted chain together with per-step evidence, transition probabilities, and a plain-language interpretation suitable for human inspection.

\subsection{\pheno{} construction}\label{sec:methods_phenograph}

To build the interaction graph, we process the corpus in four stages: segmentation, mention detection, edge orientation, and aggregation. Each published article is first divided into paragraph-sized units, which serve as the basic units of evidence throughout the framework. Within each paragraph, we identify mentions of phenomena from the controlled vocabulary using dictionary matching against canonical names and known synonyms (for example, both ``Knudsen flow'' and ``free-molecular diffusion'' map to the same canonical entry). This produces, for each paragraph~$e$, a set of mentioned phenomena $S(e)\subseteq \mathcal{V}$.

When two phenomena co-occur in the same paragraph, we treat this as a candidate interaction. Co-occurrence alone, however, does not tell us which phenomenon influences which. Edge orientation is therefore determined per-paragraph as part of the extraction step described below: in the LLM backend the language model is instructed to emit explicitly directed cause--effect tuples (using textual cues such as ``leads to,'' ``causes,'' ``results in,'' or temporal precedence), and in the fallback regex backend orientation is recovered by matching three families of discourse patterns; forward (``X leads to Y''), reverse (``X is caused by Y''), and adjunct (``due to X, Y''); against the candidate paragraph. We additionally rely on equation patterns and regime-determining relationships, but only insofar as they appear within paragraphs that the extractor processes; we do not implement an independent orientation reasoner on top.

When orientation cues are ambiguous or conflicting within a single paragraph, the paragraph is not forced into a directed relation by a fixed rule priority. Instead, the two possible directions, $u\rightarrow v$ and $v\rightarrow u$, are scored separately using the same backend. Directional support is assigned to the edge with stronger evidence, while paragraphs with insufficient directional separation contribute little or no directional confidence. This prevents simple co-occurrence from being converted into an artificially confident mechanistic direction.

In practice, candidate edges are extracted from each paragraph using one of two complementary backends. The primary backend is a large language model: each paragraph is sent to Anthropic's Claude~3.7~Sonnet (\texttt{claude-3-7-sonnet-latest}) with a strict-JSON system prompt asking for a list of directed cause--effect relations and a per-relation confidence value $c(u,v;e)\in[0,1]$. Extractions are cached per (paper, chunk) pair in a local SQLite database (\texttt{cache.sqlite}, released alongside the paper), so the released graph is fully reproducible from the cache without re-spending API credits. A fallback regex backend is also implemented and is used when API access is not available; it identifies discourse cues (``leads to,'' ``is caused by,'' ``due to,'' and similar) within sentences and assigns a hard-coded confidence of $c=0.55$ (or $c=0.65$ when the sentence contains explicit causal language). The released \pheno{} graph was built entirely with the LLM backend; the regex backend serves as an offline reproducibility check. We treat the LLM extractor as a structured-relation labeller, not as a reasoner: its outputs are deterministically aggregated and verified against corpus evidence by the downstream pipeline.

The per-paragraph confidences supporting the same directed edge are then aggregated. Let $\mathcal{P}_{u,v}$ denote the set of distinct papers from which $(u,v)$ was extracted, and let $\mathcal{E}_{u,v}$ denote the set of supporting paragraphs. Two summary statistics are stored per edge:
\begin{equation}
\mathrm{support}(u,v)=|\mathcal{P}_{u,v}|, \qquad \mathrm{conf}(u,v)=\sum_{e\in \mathcal{E}_{u,v}} c(u,v;e).
\end{equation}
The edge weight used by the downstream pipeline is the simple sum
\begin{equation}
w(u,v)=\mathrm{support}(u,v) + \mathrm{conf}(u,v),
\end{equation}
which combines paper-level diversity (how many distinct papers attest the relation) with paragraph-level evidence accumulation (how strongly each paragraph supports it). For each edge we additionally retain the top three most informative evidence paragraphs verbatim, for use in the audit report (Sec.~\ref{sec:methods_audit}). To make edge weights comparable across nodes, outgoing weights at each node are normalised:
\begin{equation}
p(u\rightarrow v)=\frac{w(u,v)}{\sum_{v'} w(u,v')}.
\end{equation}
We refer to $p(u\rightarrow v)$ as a transition probability throughout, but emphasise that it denotes a graph-normalised evidence score over outgoing edges, \emph{not} a physical transition rate or an intervention-defined causal probability: it is computed by normalising aggregated graph-edge weights and represents a model-internal evidence score derived from corpus statistics.

We use $-\log p(u\rightarrow v)$ as an evidence-weighted edge cost, so that poorly supported transitions incur a higher cost. For a multi-step reasoning chain $y=(u_1,\dots,u_T)$, we report a chain-level score equal to the mean per-step negative log-likelihood:
\begin{equation}
\DL(y)=\frac{1}{T-1}\sum_{i=1}^{T-1} -\log p(u_i\rightarrow u_{i+1}),
\end{equation}
which is the average cost of a path under the graph's induced first-order Markov chain. A low value indicates a well-supported chain; a high value flags steps that lack corpus evidence. Individual steps are flagged for review if the edge is missing from the graph, the transition score falls below a threshold $\tau_p$, or the support count is below $\tau_s$. Given a start phenomenon $s$ and a target phenomenon $t$, finding the best-supported chain reduces to a shortest-path problem~\cite{dijkstra1959shortest} on the weighted directed graph.

The released \pheno{} graph contains 635 directed edges over 782 phenomena, corresponding to a graph density of approximately $1.0\times10^{-3}$. The out-degree distribution is heavy-tailed: the median is 1, the mean is 1.44, and the maximum is 40. Half of the nodes therefore have a single supported successor, which has two implications worth stating explicitly. First, the audit-report notion of ``rank among outgoing edges'' is informative only at higher-degree nodes; at unique-successor nodes the rank carries no comparative meaning. Second, the graph is sparse compared to general-purpose scientific knowledge graphs ($10^{4}$--$10^{7}$ edges, e.g., PubChemKG, SPOKE, BioKG) by deliberate construction: we accept lower recall in exchange for higher per-edge evidential conservatism, so that every retained edge is supported by at least one paragraph passing the orientation procedure. The full out-degree distribution is provided in the Supplementary Information.

\begin{figure}[!t]
  \centering
  \includegraphics[width=\linewidth]{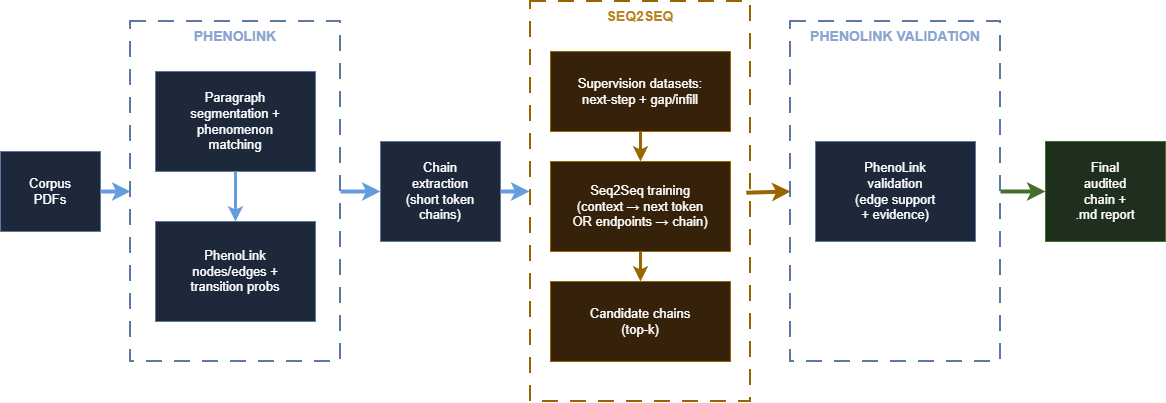}
  \caption{{\textbf{Construction of \pheno{} from a scientific corpus.} The graph supports per-step auditing of candidate chains proposed by \phenseq{} via Tier-1 evidence verification (Sec.~\ref{sec:methods_pipeline}).}}
  \label{fig:phenolink_construction}
\end{figure}

\subsection{\phenseq{}: statistical chain modelling layer}\label{sec:methods_phenoseq}

While \pheno{} captures which transitions between phenomena are supported by the literature, it cannot by itself propose new chains for unseen endpoint pairs. This is the role of \phenseq{}, a lightweight sequence model that learns the statistical patterns of how phenomena follow one another and uses these patterns to suggest plausible intermediate steps.

Training data for \phenseq{} is derived directly from \pheno{}. We extract 50{,}000 short chains (length 4--8 phenomena) by performing global random walks through the graph with top-$k=10$ next-step sampling. The walks collapse to 2{,}071 distinct trajectories: the random-walk sampler oversamples high-confidence trunks of the graph (whose median out-degree is 1), and we accept this multiplicity as a deliberate corpus-design choice. The corpus is not intended to provide trajectory diversity; it is intended to provide dense local-transition coverage so that the proposer learns well-conditioned conditional distributions over admissible successors. The walk-length distribution is heavy on short walks: 33{,}845 walks of length 4, 13{,}846 of length 5, 2{,}127 of length 6, 176 of length 7, and 6 of length 8. Of the 635 directed edges in \pheno{}, 217 (34.2\%) appear as bigrams in at least one training walk; the remaining 418 edges are reachable to the pipeline only through graph-first routing during inference. From the 50{,}000 walks we construct two supervised tasks. In the \emph{next-step prediction} task (118{,}652 examples, one for each $(u_i, u_{i+1})$ bigram in any training walk where $i\geq 2$), the model sees a prefix of two or more consecutive phenomena and predicts the next. In the \emph{fill-in-the-middle} task (52{,}309 examples, of which 50{,}000 are fixed-position masks and 2{,}309 are variable-span masks where chain length permitted), one or more phenomena in a chain are masked and the model reconstructs the complete sequence. The second task is particularly relevant for endpoint bridging: given a start and end phenomenon with a gap between them, the model proposes plausible intermediates.

We frame \phenseq{}'s effective generative vocabulary as 204 phenomena; the tokens present in the training walks, rather than the full 782-phenomenon vocabulary of \pheno{}. The remaining 578 vocabulary tokens are accessible to the pipeline only through graph-first routing (Sec.~\ref{sec:methods_pipeline}) over pre-existing edges, not through \phenseq{} generation. We return to this scope condition in Sec.~\ref{sec:results_correct} when interpreting the rare-token failure mode.

The model uses an encoder--decoder architecture~\cite{sutskever2014seq2seq} with attention~\cite{vaswani2017attention}. During inference, \phenseq{} generates candidate next steps by sampling from its learned distribution $q_\theta(v \mid u_1,\dots,u_{t})$, conditioned on the full preceding chain prefix via the Transformer encoder (self-attention over up to 64 context tokens) rather than on a fixed-length window. These candidates are then filtered against \pheno{}: only steps that correspond to edges in the graph with sufficient evidence support are retained. Beam search~\cite{freitag2017beam} explores multiple candidate chains in parallel, and the final output is the chain with the lowest combined evidence-weighted edge cost.

We disclose two overlap statistics between the gold-suite evaluation queries and these training walks, addressing the reviewer concern about training/evaluation circularity. First, of 36 gold-suite queries, 31 (86.1\%) have endpoint pairs $(A,B)$ that do not appear as start/end of any training walk: \phenseq{} is therefore not recalling memorised end-to-end chains. Second, however, 97.2\% of consecutive $(u,v)$ bigrams in expected gold-suite chains do appear in at least one training walk; only 2 of 36 expected chains have any bigram that the proposer has never seen. We frame this as the intended behaviour of the system: the random-walk corpus is designed to give the proposer dense local-transition coverage, after which the proposal/verification pipeline composes those locally-supported transitions into endpoint-to-endpoint paths the training set does not contain. The 86\% novel-endpoint-pair statistic shows this composition is non-trivial; the 97\% bigram-coverage statistic shows the composition is bounded by the graph's edge structure rather than by free-form generation.

Computationally, \phenseq{} is therefore best understood as a constrained sequence model over a finite scientific vocabulary. The decoder proposes candidate phenomenon tokens according to the learned conditional distribution, but the admissible output space is restricted by the \pheno{} graph during inference. This converts unconstrained sequence generation into graph-constrained decoding: candidate steps that do not correspond to evidence-backed edges are masked out or rejected by the verifier. The architecture therefore separates statistical proposal from evidential validation. \phenseq{} is responsible for exploring plausible intermediate phenomena whereas \pheno{} imposes hard constraints derived from corpus evidence. This separation reduces hallucinated transitions and makes every accepted chain under Tiers 0/1/2 reproducible from fixed graph artifacts. We evaluate the contribution of the \phenseq{} architecture itself, relative to a 1st-order Markov chain trained on the same walks, in Sec.~\ref{sec:results_ablation}; that ablation finds the two proposers comparable on pass rate, supporting the framing that the contribution of the framework is the proposal/verification \emph{separation} rather than the seq2seq architecture per se.

\subsection{Agentic pipeline and verification protocol}\label{sec:methods_pipeline}

The complete pipeline operates as an agentic loop that generates, evaluates, and repairs candidate chains until an acceptable explanation is found or a failure is reported. Figure~\ref{fig:pipeline_overview} illustrates the full workflow.

When a user submits a query specifying a start phenomenon $A$ and a goal phenomenon $B$, the system first attempts \emph{graph-first routing}: it checks whether an evidence-supported path from $A$ to $B$ already exists in \pheno{}. If so, this path is returned directly. If not, \phenseq{} generates a candidate chain using masked decoding, in which at each step the set of possible next tokens is restricted to phenomena that are reachable from the current position in the graph:
\[
\tilde{P}(v \mid \cdot) \propto
\begin{cases}
P(v \mid \cdot), & v \in \mathcal{A}(u) \\
0, & \text{otherwise},
\end{cases}
\]
where $\mathcal{A}(u)$ is the set of admissible successors of $u$ in the graph.

Each candidate chain is then evaluated by a tiered verification protocol with increasing stringency:

\emph{Tier~0 (structural validity)} checks that the output is non-empty, correctly formatted, and that the chain begins at $A$ and ends at $B$. Violations such as endpoint mismatches or repetition loops are rejected immediately.

\emph{Tier~1 (evidence verification)} examines each step $u\rightarrow v$ in the chain against \pheno{}. For the chain to pass, every step must correspond to an edge in the graph with support and weight exceeding predefined thresholds. Tier~1 outputs include a per-edge summary (support count, weight, and number of evidence paragraphs) and a chain-level \emph{Total Evidence Score} (TES), computed as the sum of all edge weights along the chain. The TES provides a single number summarising how well the entire explanation is supported by the literature.

\emph{Tier~2 (semantic plausibility)} applies a deterministic filter to catch physics-level inconsistencies that are structurally valid but semantically nonsensical; for example, a chain that links two phenomena through an intermediate that belongs to an unrelated physical domain.

\emph{Tier~3}, when enabled, invokes an external language model to assess whether the proposed chain is physically plausible beyond what graph structure alone can verify. The referee produces a plausibility score between 0 and 1, optional per-edge rationales, and suggested waypoint phenomena that might improve a failed chain. Three referee backends were evaluated: an open-source model (Qwen2.5-14B-Instruct via vLLM), OpenAI's GPT-4o-mini (\texttt{gpt-4o-mini-2024-07-18}) via the OpenAI Chat Completions API, and Anthropic's Claude~Opus~4.7 (\texttt{claude-opus-4-7}) via the Anthropic Messages API. The Claude Opus~4.7 referee is invoked at \mbox{$\mathrm{temperature}{=}0$} (the field is omitted from the request payload, as it is deprecated for this model class), with a strict-JSON system prompt; transient HTTP errors and model refusals are surfaced distinctly in the released audit logs (\textsc{referee\_error} versus \textsc{referee\_refusal}, the latter carrying the upstream \texttt{request\_id}) so that the fraction of chains affected by either condition is auditable. In the runs reported below, the Claude Opus~4.7 referee returned a model refusal on 2 of 36 gold-suite queries and 3 of 66 stress-suite queries; in all cases the underlying chain had already been classified by Tiers~0--2 and the refusal did not flip a previously-passed verdict to a hallucinated one.

We deliberately use a different Claude generation for the Tier-3 referee than was used to extract the graph: graph extraction was performed with Claude~3.7~Sonnet (Sec.~\ref{sec:methods_phenograph}), while the Tier-3 referee runs against Claude~Opus~4.7. The two models are separated by multiple model generations, distinct training runs, and different parameter scales, which mitigates; though does not eliminate the concern that the referee could simply rubber-stamp the extractor's choices. We additionally report Qwen and GPT-4o-mini referees as cross-vendor controls.

We emphasise that Tier-3 accept/reject decisions and waypoint suggestions are produced by an opaque language-model judgment and are \emph{not} themselves auditable in the corpus-evidence sense: only the per-step scores and supporting paragraphs returned by Tiers~0--2 satisfy the auditability claim made elsewhere in this paper. We therefore present Tier-3 as an \emph{optional ablation} to the core graph-backed pipeline rather than as a first-class configuration. As shown in Sec.~\ref{sec:results}, enabling Tier-3 in fact \emph{reduces} both gold-suite pass rate and stress-suite correct-behaviour relative to the Tier~0/1/2 baseline; we retain Tier-3 in the evaluation to characterise where LLM plausibility judgments add or subtract value rather than to claim it as part of the auditable core.
If a candidate chain fails at any tier, the system executes \emph{repair actions} within a bounded retry budget: increasing beam width for broader exploration, splitting the problem into two subproblems ($A \rightarrow W$ and $W \rightarrow B$) using referee-suggested waypoints, or falling back to stricter graph-only search. If all repair attempts are exhausted and the endpoints remain unreachable, the system returns an \emph{explanatory failure state} identifying the reason (e.g., disconnected components, missing evidence) rather than producing an unsupported chain.

\subsection{Evaluation design}\label{sec:methods_eval}

To evaluate the pipeline, we designed 172 test queries divided into four suites with zero overlap with any \phenseq{} training data. We define ``zero overlap'' at the supervised-example level rather than at the vocabulary level: evaluation queries may contain phenomenon tokens that also appear in \pheno{}, because the task is not open-vocabulary recognition. The complete endpoint query and its target chain were excluded from the \phenseq{} training examples. This split tests whether the pipeline can reconstruct or reject endpoint-conditioned mechanistic chains using the graph-backed evidence layer, rather than simply memorising training sequences. Bigram-level and endpoint-pair-level overlap statistics are reported in Sec.~\ref{sec:methods_phenoseq}.

The four test suites are:

\begin{itemize}
\item \textbf{Gold suite} ($n=36$): graph-consistent chains unseen in training. Each query specifies start phenomenon $A$ and end phenomenon $B$; the expected behaviour is to produce a valid evidence-supported chain $A \rightarrow \cdots \rightarrow B$.

\item \textbf{Stress suite} ($n=66$): an adversarial robustness test. For 50 of the 66 queries, the correct system behaviour is graceful rejection rather than chain production. The suite spans six categories: reversed edges, disconnected endpoints, dead-end start nodes, out-of-domain queries, rare-token queries (valid edges but zero training appearances of the start token), and self-loops.

\item \textbf{OOV suite} ($n=39$): tests robustness to out-of-vocabulary inputs and aliases. Includes \emph{alias\_resolve} queries (e.g., \texttt{apparent\_viscosity} for \texttt{viscosity}; expected: produce a chain after resolving), \emph{fuzzy\_resolve} queries (typo-style misspellings), and \emph{unresolvable} queries (off-domain tokens such as \texttt{black\_hole}, \texttt{cryptocurrency}; expected: reject at endpoint resolution).

\item \textbf{Missing-edge suite} ($n=31$): queries deliberately constructed to test refusal capability. Endpoint pairs are chosen such that no evidence-supported path exists in \pheno{}. Expected behaviour: refuse rather than fabricate. Subcategories: \emph{missing\_direct\_edge} ($n=17$) and \emph{missing\_intermediate} ($n=14$).
\end{itemize}

The \textbf{baseline} configuration uses only Tiers~0, 1, and 2. We compare this against three referee-augmented configurations using GPT-4o-mini, Qwen, and Claude Opus 4.7. We additionally compare against a 1st-order Markov chain proposer trained on the same random-walk corpus (Sec.~\ref{sec:results_ablation}) and against vanilla Claude Opus 4.7 prompted zero-shot on the same 172 queries (Sec.~\ref{sec:results_external}).

For all configurations, we report the \emph{pass rate}, the \emph{correct-behaviour rate}, the \emph{Total Evidence Score} (TES) of accepted chains, the \emph{graph-supported transition fraction} (fraction of consecutive $(u,v)$ steps in produced chains that correspond to edges in \pheno{}), and the mean chain length. Confidence intervals were computed by nonparametric bootstrap resampling~\cite{efron1979bootstrap} (10{,}000 samples). Pairwise differences were assessed using two-proportion $z$-tests with Holm--Bonferroni correction; for cells with low counts ($<5$ in the failure cell), Fisher's exact test was used as a robustness check. Full pairwise statistics are reported in the released audit logs.

\subsection{Audit report generation}\label{sec:methods_audit}

For every query, regardless of outcome, the system produces a human-readable audit report as a Markdown file. Each report contains the original query in both natural-language and token form, the full predicted chain with resolved phenomenon names, and a step-by-step justification table. For each step $u \rightarrow v$, the table lists the graph transition probability $p(u \rightarrow v)$ and its rank among all outgoing edges from $u$, the support count and aggregated confidence, up to three supporting corpus paragraphs verbatim, and the top alternative successors from the same source node. Unlike purely generative explanations, these reports provide a compact, reproducible reasoning trace in which every predicted step is explicitly auditable.

\section{Results}\label{sec:results}

\begin{figure*}[!t]
    \centering
    \includegraphics[width=\textwidth]{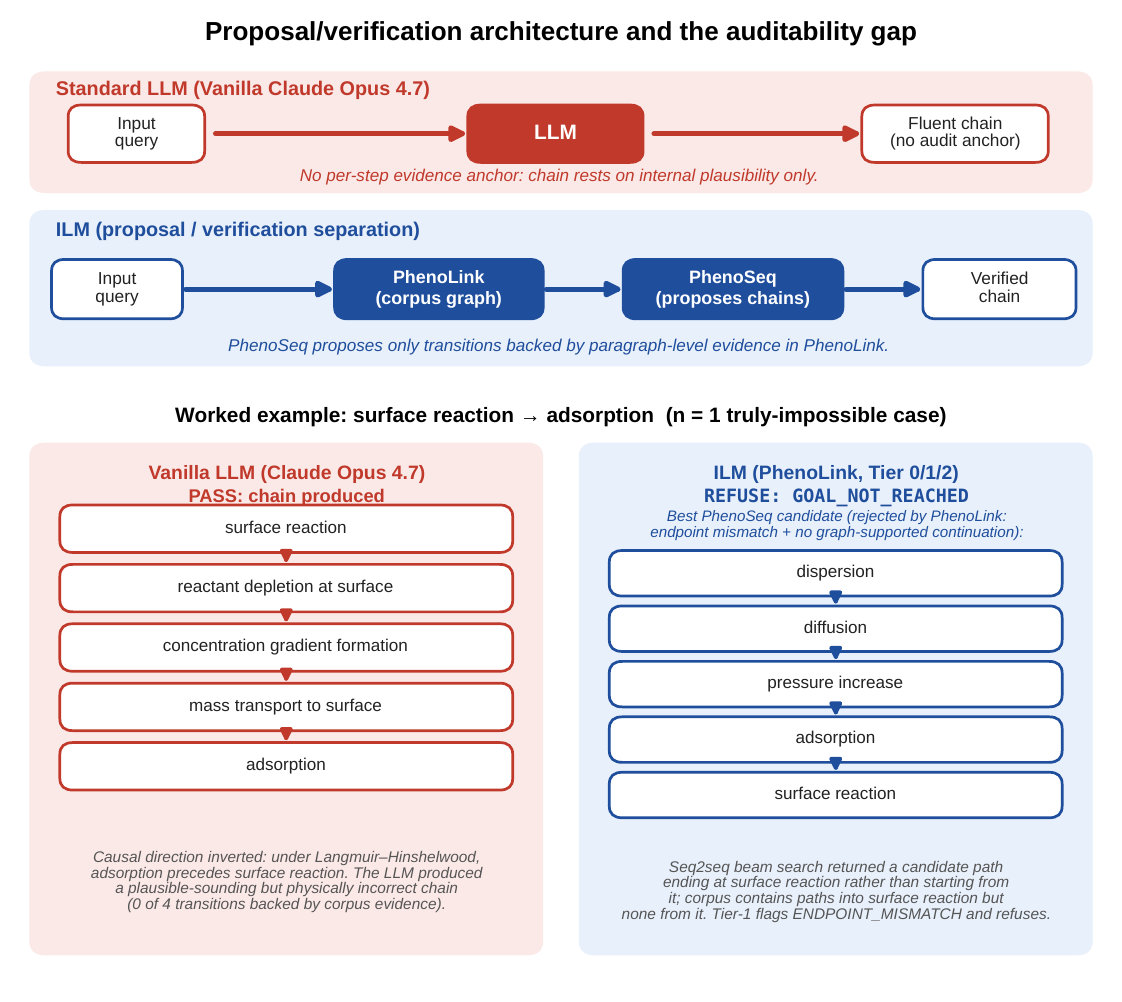}
    \caption{Proposal/verification architecture and the auditability gap.
(a) The Interaction Language Model framework separates statistical chain proposal
(PhenoSeq, a closed-vocabulary sequence model over canonical phenomena) from
corpus-grounded verification (PhenoLink, a precompiled directed graph in which
every accepted transition $u \to v$ is backed by paragraph-level evidence with
a graph-normalised score). A standard LLM, in contrast, produces fluent chains
with no per-step audit anchor. The contrast here is about per-transition
auditability rather than chain-level correctness, vanilla Claude Opus 4.7
and ILM are comparable on the gold-suite pass rate (91.7\% vs 94.4\%).
(b) Representative truly-impossible missing-edge query (\texttt{surface\_reaction} $\to$
\texttt{adsorption}). Vanilla Claude Opus 4.7 produces a 4-step
chain that is fluent but reverses the canonical Langmuir--Hinshelwood causal
direction; the PhenoLink Tier-0/1/2 pipeline refuses with \textsc{goal\_not\_reached}.
Across all 146 chains that vanilla Claude Opus 4.7 produced across the four
evaluation suites (gold, stress, out-of-vocabulary, missing-edge), only 4 of
568 consecutive transitions correspond to edges in PhenoLink (0.70\%);
PhenoLink chains, by construction, achieve 100\%. This 0.70\% vs 100\%
auditability gap is structural and does not depend on refusal-rate claims.}
    \label{fig:arch_audit}
\end{figure*}

Figure~\ref{fig:arch_audit} contrasts the ILM proposal/verification architecture with a standard LLM, illustrating both the high-level design and a concrete worked example. The top panel shows that the ILM framework routes input queries through \pheno{} (the corpus-grounded directed graph) and \phenseq{} (the closed-vocabulary sequence proposer), so every accepted transition is anchored to paragraph-level evidence; a standard LLM, by contrast, produces fluent chains directly from query to output with no per-step audit anchor. The bottom-panel worked example (\texttt{surface\_reaction} $\to$ \texttt{adsorption}) demonstrates the operational consequence: vanilla Claude Opus 4.7 produces a 4-step chain that reverses the canonical Langmuir--Hinshelwood causal direction, while the ILM Tier-0/1/2 pipeline refuses the same query because no graph path connects the endpoints. Across all 146 chains that vanilla Claude Opus 4.7 produced on our 172-query evaluation suite, only 4 of 568 consecutive transitions (0.70\%) correspond to edges in \pheno{}; ILM chains, by construction, achieve 100\%. This per-transition auditability gap, rather than aggregate pass-rate superiority on standard queries, is the operational distinction we frame as the core contribution of the framework.

\subsection{Pipeline performance on gold and stress suites}\label{sec:results_overall}

We evaluated the \pheno{}--\phenseq{} pipeline on two predefined test suites with zero training overlap: a gold suite of in-distribution endpoint queries ($n=36$) and an adversarial stress suite ($n=66$). The stress suite was designed so that for 50 of its 66 queries, the correct system behaviour is graceful rejection rather than chain production; it encompasses reversed edges, disconnected graph components, dead-end start nodes, out-of-domain phenomena, rare tokens, and self-loops (Table~\ref{tab:suites}).

\begin{figure}[htbp]
    \centering
    \includegraphics[width=\linewidth]{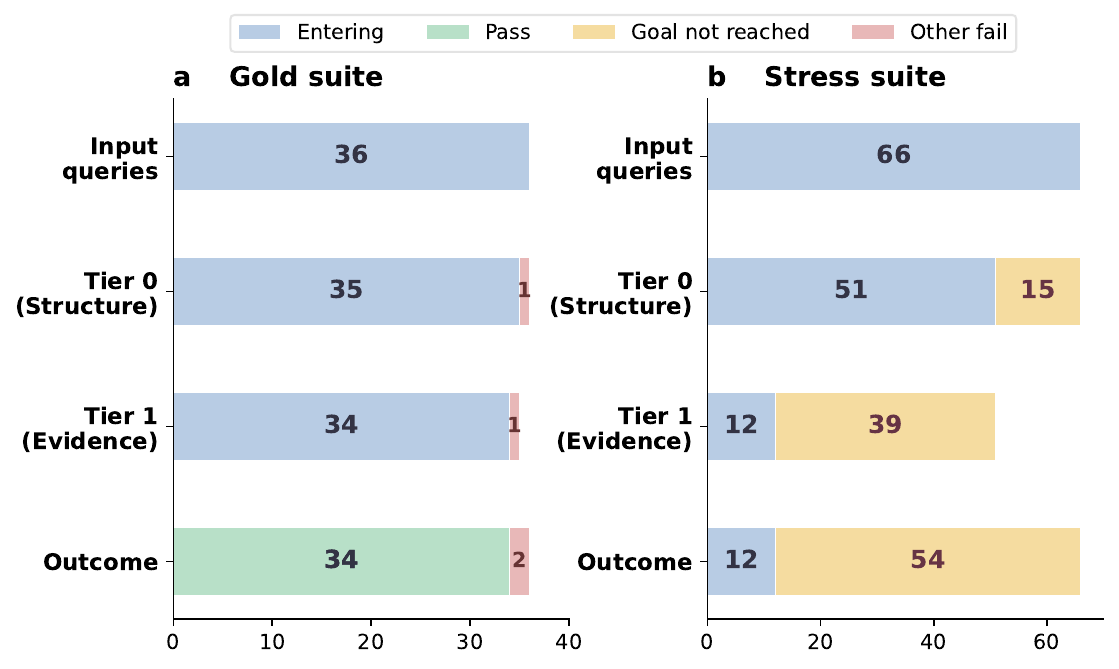}
    \caption{Query flow through the verification tiers for the Tier~0/1/2 baseline on both suites. The gold suite retains most queries through all tiers, while the stress suite experiences substantial attrition at Tier~1 (evidence verification).}
    \label{fig:sankey}
\end{figure}

The pipeline's core task is to take two endpoint phenomena (a cause and an effect) and construct a coherent mechanistic chain connecting them, with every intermediate step grounded in corpus evidence. For example, given the query \texttt{temperature\_gradient [GAP] mass\_transfer}, the system must determine that a temperature gradient drives diffusion (supported by 12 corpus paragraphs describing temperature-gradient-dependent diffusion in contexts ranging from methane transport in coal to CO$_2$ permeation in porous liquids) and that diffusion in turn drives mass transfer (supported by 5 paragraphs linking diffusion coefficients to mass flux). The resulting chain temperature\_gradient $\rightarrow$ diffusion $\rightarrow$ mass\_transfer is not merely a statistical correlation but a mechanistic pathway in which each step is backed by specific physical evidence and assigned a quantitative transition probability. Equally important, when asked to construct a chain in a physically implausible direction (for instance, clustering $\rightarrow$ mass\_transfer), the system must recognise that no corpus evidence supports this direction and refuse to produce a chain rather than hallucinate one. These two capabilities, constructing evidence-grounded chains and rejecting unsupported ones, are what we evaluate across the gold and stress suites. To assess these capabilities systematically, we report three complementary views of system behavior: aggregate pass and correct-behavior rates (Table~\ref{tab:main}), a per-category breakdown of stress responses (Table~\ref{tab:stress_correct}; Fig.~\ref{fig:correct_behaviour}), and the flow of queries through the verification tiers (Fig.~\ref{fig:sankey}). We discuss each in turn.

Table~\ref{tab:main} reports pass rate, correct-behaviour rate, mean Total Evidence Score (TES), and mean chain length for the Tier-0/1/2 baseline and three referee-augmented variants. On the gold suite, the Tier-0/1/2 baseline achieves a 94.4\% pass rate (95\% bootstrap CI 86--100\%), matching or exceeding all three referee-augmented configurations (80.6--88.9\%). On the stress suite, raw pass rates are deliberately low across configurations (18.2--28.8\%) because 50 of the 66 stress queries should be rejected rather than answered; the more informative metric is correct-behaviour rate, on which the baseline again leads (69.7\% versus 59.1--68.2\% for referee variants). The Total Evidence Score column shows that pass-rate parity is not bought by accepting unsupported chains: gold-suite accepted chains carry TES 9.4--10.2 across configurations, indicating consistent evidential depth. Two findings deserve emphasis: (i) the graph-backed verification stack alone is sufficient for in-distribution mechanistic reasoning, and (ii) adding an LLM referee shifts but does not improve aggregate performance, the referees override the structural self-loop signal (Sec.~\ref{sec:results_correct}) and occasionally reject graph-supported chains as physically implausible, both of which reduce correct-behaviour rate.

Table~\ref{tab:stress_correct} decomposes the stress suite into six adversarial categories and reports correct-behaviour rate by category and configuration. All four configurations achieve 100\% correct rejection of reversed-edge queries: this is a structural property of the directed graph (a reversed edge has no valid path) rather than a learned capability. Self-loop queries show the largest configuration-level divergence: the Tier-0/1/2 baseline detects all 8 self-loops (the trivial single-token chain $\{A\}$ vacuously satisfies the verification tiers), but every referee-augmented configuration drops to 0/8 because the Tier-3 referee accepts $A \to A$ as physically plausible identity, overriding the baseline's correct-behaviour verdict. The Claude Opus 4.7 referee dominates dead-end queries (10/10 versus 6--7 for other configurations) and matches the strongest configurations on disconnected and out-of-domain queries. The sole universal failure is the rare-token category (0/8 across all configurations), where the seq2seq proposer cannot emit tokens with zero training appearances: a bound we discuss further in Sec.~\ref{sec:results_failures}.

Figure~\ref{fig:sankey} traces the flow of queries through the verification tiers for the Tier-0/1/2 baseline. On the gold suite (panel a), 35 of 36 queries pass Tier 0, 34 pass Tier 1, and 34 are accepted; the single Tier-0 failure is a malformed endpoint structure rather than a corpus-evidence problem. On the stress suite (panel b), attrition occurs almost entirely at Tier 1: of 51 queries that pass Tier 0, only 12 survive evidence verification, and 39 are correctly rejected as \textsc{goal\_not\_reached} because no graph-supported path connects their endpoints. The figure makes the central design property visible: stress-suite attrition is not model failure but correct refusal driven by the graph structure, and the verification tiers act as a sieve that separates queries answerable from corpus evidence from queries that require fabrication.

The Tier~0/1/2 baseline without any LLM referee achieved a 94.4\% pass rate [95\% CI: 86--100\%] on the gold suite (Table~\ref{tab:main}), confirming that the graph-backed core pipeline is sufficient for in-distribution mechanistic reasoning. Stress-suite pass rates were lower (18--29\% across configurations), but this reflects the suite's adversarial design: when measured by \emph{correct-behaviour rate} (correctly rejecting invalid queries or producing valid chains as appropriate), all configurations achieved 70--80\% correct responses (Table~\ref{tab:stress_correct}; Fig.~\ref{fig:correct_behaviour}).

Table~\ref{tab:suites} summarises the evaluation design: 172 queries divided into four held-out suites with zero supervised-example overlap with \phenseq{} training. The gold suite ($n{=}36$) tests in-distribution chain reconstruction; the stress suite ($n{=}66$) is the adversarial robustness test with six subcategories spanning reversed edges, disconnected endpoints, dead-end starts, out-of-domain phenomena, rare tokens, and self-loops. The OOV suite ($n{=}39$) tests robustness to vocabulary edge cases (alias resolution, fuzzy matching, off-domain refusal), and the missing-edge suite ($n{=}31$) tests refusal capability by constructing endpoint pairs with no supported path. Crucially, 50 of the 66 stress queries and all 31 missing-edge queries are designed such that the correct system behaviour is graceful rejection rather than chain production--the suites jointly evaluate both productive accuracy and refusal accuracy. All four suites and their per-query expected behaviours are released as \texttt{test\_*.jsonl} artifacts to enable independent reproduction.

\begin{table}[htbp]
\centering
\caption{Complete evaluation map across the four held-out suites. Each query specifies expected behaviour and the source from which it was constructed. Released artifact filenames are given in the last column.}
\label{tab:suites}
\setlength{\tabcolsep}{4pt}
\footnotesize
\begin{tabular}{llp{2.5cm}llp{2.5cm}}
\toprule
\textbf{Suite} & $n$ & \textbf{Subcategory} & \textbf{Expected} & \textbf{Source} & \textbf{Artifact} \\
\midrule
Gold & 36 & graph-consistent & PASS & held-out edges, unseen at training & \texttt{test\_gold.jsonl} \\
\addlinespace
\multirow{6}{*}{Stress} & 15 & reversed edges & Reject & forward edge in graph; reverse queried & \multirow{6}{2.5cm}{\texttt{test\_stress.jsonl}} \\
& 15 & disconnected & Reject & endpoints in distinct components & \\
& 10 & dead-end start & Reject & start node has zero outgoing edges & \\
& 10 & out-of-domain & Reject/flag & non-diffusion tokens & \\
& 8 & rare token & PASS & valid 1-hop edge, zero training appearances & \\
& 8 & self-loop & Detect & $A = B$ & \\
\addlinespace
\multirow{3}{*}{OOV} & 14 & alias\_resolve & PASS & synonym of in-vocab token & \multirow{3}{2.5cm}{\texttt{test\_oov.jsonl}} \\
& 15 & fuzzy\_resolve & PASS & typo of in-vocab token & \\
& 10 & unresolvable & Reject & off-domain token & \\
\addlinespace
\multirow{2}{*}{Missing-edge} & 17 & missing\_direct\_edge & Reject & endpoint pair, no edge in graph & \multirow{2}{2.5cm}{\texttt{test\_missing\_edge.jsonl}} \\
& 14 & missing\_intermediate & Reject & no supported path & \\
\bottomrule
\end{tabular}
\end{table}

\begin{table}[htbp]
\centering
\caption{Overall performance grouped by configuration. Cells report count/total (percentage). Model identifiers: GPT-4o-mini = \texttt{gpt-4o-mini-2024-07-18} (OpenAI Chat Completions API); Qwen = \texttt{Qwen2.5-14B-Instruct} via vLLM; Claude Opus 4.7 = \texttt{claude-opus-4-7} (Anthropic Messages API). API calls use $\mathrm{temperature}{=}0$ where applicable (the field is omitted from request payloads for \texttt{claude-opus-4-7}, as it is deprecated for this model class). Beam width 8, max retries 2, semantic evaluation enabled, identical query order across configurations. Random seeds documented in the released audit logs.}
\label{tab:main}
\setlength{\tabcolsep}{4pt}
\footnotesize
\begin{tabular}{lccccc}
\toprule
\textbf{Configuration} & \textbf{Gold pass} & \textbf{Gold TES} & \textbf{Stress pass} & \textbf{Stress correct-beh.} & \textbf{Mean chain len} \\
\midrule
PhenoLink Tier 0/1/2 (baseline) & 34/36 (94.4\%) & 9.9 & 12/66 (18.2\%) & 46/66 (69.7\%) & 2.15 \\
PhenoLink + GPT-4o-mini referee & 32/36 (88.9\%) & 9.4 & 17/66 (25.8\%) & 41/66 (62.1\%) & 2.16 \\
PhenoLink + Qwen referee & 31/36 (86.1\%) & 8.4 & 19/66 (28.8\%) & 39/66 (59.1\%) & 2.10 \\
PhenoLink + Claude Opus 4.7 referee & 29/36 (80.6\%) & 10.0 & 13/66 (19.7\%) & 45/66 (68.2\%) & 2.15 \\
\addlinespace
Markov 1st-order (ablation) & 33/36 (91.7\%) & 10.16 & 18/66 (27.3\%) & 56/66 (84.8\%) & 2.21 \\
\addlinespace
Vanilla Claude Opus 4.7 (zero-shot) & 33/36 (91.7\%) & -- & 7/66 (10.6\%) & 25/66 (37.9\%) & 3.82 \\
\bottomrule
\end{tabular}
\end{table}

\subsection{Adversarial robustness by stress category}\label{sec:results_correct}

\begin{figure}[htbp]
    \centering
    \includegraphics[width=\linewidth]{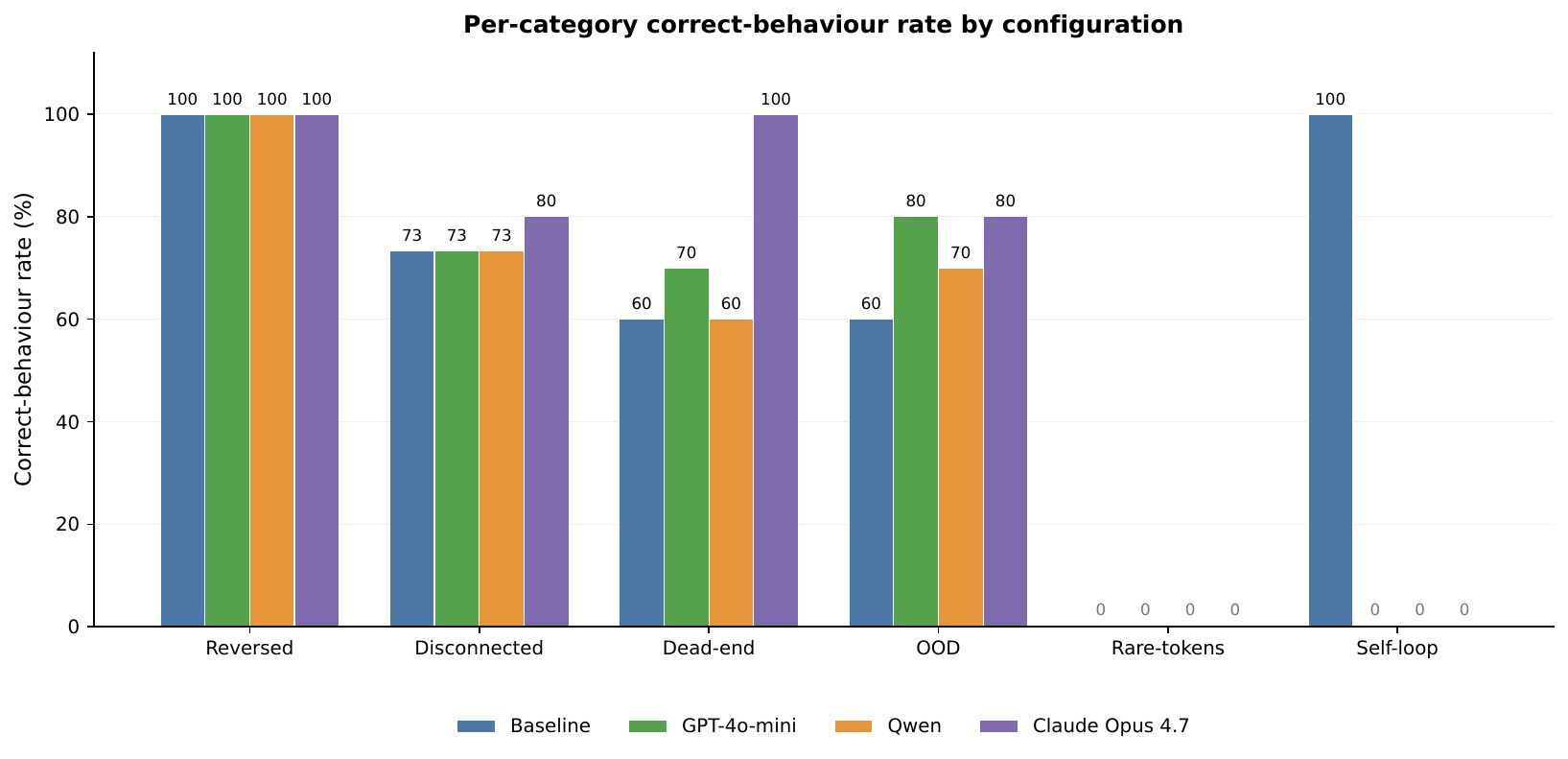}
    \label{fig:correct_behaviour}

\caption{Per-category correct-behaviour rate by configuration on the stress suite. All four configurations achieve 100\% correct rejection of reversed-edge queries (caught structurally by the directed graph). The Tier-0/1/2 baseline (no referee) also achieves 100\% on self-loop detection; referee variants reduce this to 0\% by accepting the trivial $A \rightarrow A$ chain. The Claude Opus 4.7 referee dominates dead-end queries (100\% vs.\ 60--70\% for others). Rare-token category fails at 0\% across all configurations because the seq2seq proposer has zero training appearances of the start token (Sec.~\ref{sec:results_correct}). Numerical values are reproduced from the recomputed correct-behaviour artifact released alongside the paper.}

\end{figure}

All configurations achieved 100\% correct rejection of reversed-edge queries and 100\% correct detection of self-loops, demonstrating that the graph-backed verification layer reliably enforces edge directionality (Table~\ref{tab:stress_correct}; Fig.~\ref{fig:correct_behaviour}).

For instance, the query \texttt{dispersion [GAP] mass\_transfer} is correctly rejected by every configuration. The corpus supports a single direction, \texttt{mass\_transfer} $\to$ \texttt{dispersion} (weight 1.9, one supporting paragraph from chromatographic transport literature), and no evidence for the reverse. The pipeline fails at Tier 1 with \textsc{goal\_not\_reached}, demonstrating that edge directionality is enforced structurally by the graph rather than learned probabilistically.

Stress-suite composition analysis reveals that aggregate improvements from referee-enabled configurations were partly driven by resolution of 8 endpoint-identical (self-loop) queries, trivially handled by the referee but unreachable by the baseline decoder. On the harder endpoint-distinct subset ($n=58$), the graph-backed baseline remained competitive (20.7\%), and improvements from referee-guided repair were limited by a chain-completion bottleneck in weakly connected graph regions.

 All configurations correctly handled reversed-edge queries (Table~\ref{tab:stress_correct}). For reversed-edge queries the rejection is structural: a reversed edge has no valid path in the directed graph and is rejected by Tier-1 evidence verification. Self-loops are similarly handled structurally by the Tier-0/1/2 baseline (8/8 detected via the trivial single-token chain $\{A\}$ that vacuously passes Tiers~0--2 under our rubric). However, every referee-augmented configuration reduces self-loop correct-behaviour to 0/8: a Tier-3 LLM referee, asked whether the trivial chain $A \rightarrow A$ is plausible, accepts it (correctly recognising identity), but our rubric counts that PASS verdict as incorrect because the desired system output for self-loops is detection rather than acceptance. We retain self-loop in the suite as a transparency check on the rubric's interaction with the referee rather than as a learned-reasoning evaluation. The same caveat applies to disconnected-component queries, which are caught by graph reachability rather than by any model judgment.

The Baseline (Tier 0/1/2 with no referee) achieves the highest overall correct-behaviour rate at 69.7\% (46/66), with the Claude Opus 4.7 referee a close second at 68.2\% (45/66). Adding any referee variant reduces correct-behaviour relative to the no-referee baseline because referees override the structural self-loop FAIL signal. Within referee variants, the Claude Opus 4.7 configuration dominates on dead-end queries (10/10 correct refusals vs.\ 6--7 across other configurations) and matches the strongest other configurations on disconnected (12/15) and out-of-domain (8/10). We read the configuration differences as referee-calibration sensitivity: GPT-4o-mini is best on out-of-domain rejection but does not help on dead-end queries, while Claude Opus 4.7 is best on dead-end and disconnected but neutral elsewhere.

All configurations failed entirely on the rare-token category (0/8), where the seq2seq proposer has zero training appearances of the start token. Although a direct 1-hop edge with $p=1.0$ exists from the rare-token start to its goal in every case, the proposer fails to emit that edge and produces a multi-hop chain that does not reach the requested goal; Tier-1 correctly flags this as GOAL\_NOT\_REACHED. None of the Tier-3 referees recover the failure, because referees suggest waypoint phenomena rather than collapsing the chain to the missing direct edge. This bounds the effective generative scope of \phenseq{} to the 204 phenomena present in the training walks; the remaining ${\sim}74\%$ of the declared vocabulary is reachable only via graph-first routing through pre-existing edges. We frame this not as a marginal limitation but as a defining scope boundary of the current \phenseq{} layer. Densifying training-walk coverage of the graph, or adopting an architecture that does not require token-level training presence (for example, a graph neural network), is the natural remedy and is left for future work.


\begin{table}[htbp]
\centering
\caption{Stress-suite correct-behaviour by adversarial category. Cells show count/total (percentage); 95\% Wilson binomial confidence intervals shown for the four smallest categories. ``Detect'' (self-loop row): the system returns a single-token chain $\{A\}$ when start equals end; the rubric counts this as correct for the Tier-0/1/2 baseline (vacuous Tier-0/1/2 pass) but as incorrect for referee-augmented configurations (the referee accepts the trivial chain as a multi-step pass, overriding the structural FAIL signal).}
\label{tab:stress_correct}
\setlength{\tabcolsep}{6pt}
\footnotesize
\begin{tabular}{lcccccc}
\toprule
\textbf{Category} & $n$ & \textbf{Expected} & \textbf{Baseline} & \textbf{GPT-4o-mini} & \textbf{Qwen} & \textbf{Claude Opus 4.7} \\
\midrule
Reversed & 15 & Reject & 15/15 (100\%) & 15/15 (100\%) & 15/15 (100\%) & 15/15 (100\%) \\
Disconnected & 15 & Reject & 11/15 (73\%) & 11/15 (73\%) & 11/15 (73\%) & 12/15 (80\%) \\
Dead-end & 10 & Reject & 6/10 (60\% [27,86]) & 7/10 (70\% [35,93]) & 6/10 (60\% [27,86]) & 10/10 (100\% [69,100]) \\
OOD & 10 & Reject & 6/10 (60\% [27,86]) & 8/10 (80\% [44,97]) & 7/10 (70\% [35,93]) & 8/10 (80\% [44,97]) \\
Rare-token & 8 & PASS & 0/8 (0\% [0,37]) & 0/8 (0\% [0,37]) & 0/8 (0\% [0,37]) & 0/8 (0\% [0,37]) \\
Self-loop & 8 & Detect & 8/8 (100\% [63,100]) & 0/8 (0\% [0,37]) & 0/8 (0\% [0,37]) & 0/8 (0\% [0,37]) \\
\midrule
\textbf{Overall} & 66 &        & \textbf{46/66 (69.7\%)} & \textbf{41/66 (62.1\%)} & \textbf{39/66 (59.1\%)} & \textbf{45/66 (68.2\%)} \\
\bottomrule
\end{tabular}
\end{table}

\subsection{Out-of-vocabulary and missing-edge robustness}\label{sec:results_oov_missing}

Beyond the 102 in-distribution and stress queries, we evaluate the pipeline on two additional held-out suites that probe failure modes specifically: 39 out-of-vocabulary queries and 31 missing-edge queries.

On the OOV suite, the Tier-0/1/2 baseline achieves a 17.9\% pass rate (95\% bootstrap CI 7.7--30.8\%) and 43.6\% correct-behaviour rate (28.2--59.0\%). Adding the Claude Opus 4.7 referee modestly changes both numbers (15.4\% pass, 41.0\% correct-behaviour); both shifts are within bootstrap CIs. The dominant OOV failure mode is alias resolution: of 14 alias\_resolve queries, the pipeline correctly maps and produces a chain for 1 in both configurations. The pipeline correctly rejects all 10 unresolvable queries (off-domain tokens such as \texttt{black\_hole}, \texttt{cryptocurrency}). Alias resolution at the current \texttt{--oov\_auto\_map\_threshold} setting (0.92) is conservative; relaxing this threshold would improve alias\_resolve at the cost of unresolvable rejection accuracy and is left as a calibration question for future work.

For example, the query \texttt{prior\_rituximab\_exposure} $\to$ \texttt{diffusion} (an oncology term paired with a diffusion phenomenon, drawn from the out-of-domain stress queries) is correctly refused. The pipeline identifies \texttt{prior\_rituximab\_exposure} as off-domain rather than substituting a lexically-close diffusion vocabulary token, and the Tier-3 referee independently assigns the lowest plausibility score in the dataset (0.08). The pipeline declines to fabricate a chain through whichever vocabulary token happens to be nearest, demonstrating conservative resolution as a deliberate design property: the same behaviour is responsible for the 100\% correct-rejection rate on the 10 out-of-domain stress queries reported in Sec.~\ref{sec:results_correct}.

On the missing-edge suite, the referee provides the strongest concrete improvement we observe: correct rejection rises from 83.9\% (Tier-0/1/2 baseline) to 93.5\% with the Claude Opus 4.7 referee. The improvement is concentrated in the missing\_intermediate subcategory (78.6\% $\rightarrow$ 100\%). Three chains that the seq2seq+Tier-0/1/2 stack accepts as PASS are correctly flagged as REFEREE\_IMPLAUSIBLE under Claude Opus 4.7's plausibility check. This is the cleanest referee win in the evaluation: graph structure alone cannot distinguish a chain that traverses weakly-connected components from a chain that traverses well-supported ones, but the referee can.

\subsection{PhenoSeq vs.\ Markov chain proposer}\label{sec:results_ablation}

To assess whether the seq2seq architecture of \phenseq{} provides capability beyond a simpler statistical baseline, we replace it with a 1st-order Markov model trained on the same 50,000 random walks. The Markov model proposes the next phenomenon based only on the current one, using bigram transition counts with add-1 smoothing. We then evaluate its proposed chains using the same Tier-0/1/2 verification stack. All hyperparameters, including beam width 8, max retries 2, and semantic evaluation, as well as the graph-admissibility mask, are kept identical. Thus, any difference in performance reflects the contribution of the \phenseq{} architecture itself rather than differences in evaluation settings or constraints.

More concretely, the Markov proposer is a $|V| \times |V|$ probability table $T$ with entries $T(v \mid u) = (n(u,v) + 1) / \sum_w (n(u,w) + 1)$, where $n(u,v)$ counts the number of times the bigram $(u,v)$ appears across the 50{,}000 training walks. Add-1 (Laplace) smoothing ensures every transition receives a non-zero probability. The model has no latent embedding space, no attention mechanism, no encoder--decoder structure, and no notion of an endpoint goal: at inference it samples the next phenomenon as a function of the current one only, with the same graph-admissibility mask filtering out non-edge candidates. Beam search and the Tier-0/1/2 verification stack are applied identically to the Markov-proposed chains and to the \phenseq{}-proposed chains.

On the gold suite, \phenseq{} achieves a 94.4\% pass rate (34/36) versus Markov's 91.7\% (33/36): a 2.7-point advantage that is not statistically significant under Holm--Bonferroni correction (Fisher's exact $p = 1.00$). \phenseq{} produces 13 chains that exactly match the expected gold chain compared to Markov's 11. On the stress suite, the two proposers are also comparable on raw pass rate (\phenseq{} 19.7\%, Markov 27.3\%) but differ markedly on correct-behaviour rate, where Markov benefits from being incapable of goal-conditioned hallucination: on 50 stress queries that should be rejected, Markov fails to produce a goal-reaching chain more often than \phenseq{}, and these failures count as correct rejections under our rubric.

To unpack two of the numbers in the previous paragraph: the `exact gold-chain match' count measures how many of the 36 gold-suite queries had their proposer-produced chain coincide \emph{token-for-token} with the expert-annotated reference chain. \phenseq{} exactly recovered the reference chain for 13 of 36 gold queries while the Markov proposer recovered it for 11; on the remaining queries both proposers reached the requested endpoints via different but evidence-supported paths, and the verification stack accepted both as valid, this is the source of the comparable aggregate pass rates. The stress-suite numbers tell a different story: when a query \emph{should} be rejected, an over-eager proposer that always finds \emph{some} path will incorrectly accept too many queries, while a less expressive proposer that frequently terminates without reaching the goal will incorrectly fail more queries, but on the adversarial stress suite this `failure' coincides with the desired behaviour. We include the Markov proposer deliberately, as a stripped-down statistical baseline that holds the verification stack fixed and isolates the contribution of proposer architecture; the auditability and refusal guarantees of PhenoLink persist in both configurations.

The fundamental difference between ILM and a Markov-chain proposer is not the architectural choice within the proposer but the \emph{framework-level separation between statistical proposal and corpus-grounded verification}. A 1st-order Markov chain proposes the next phenomenon from the current state only. A higher-order ($k$-gram) Markov chain conditions on the last $k$ states at the cost of $|V|^k$ parameters: for $|V|=782$, $k=3$ requires approximately $4.8 \times 10^8$ entries and $k=5$ requires approximately $2.9 \times 10^{14}$, infeasible to estimate from our corpus of $\sim$250{,}000 training bigrams. \phenseq{}, as an encoder--decoder with attention, conditions on the entire prefix and on the endpoint goal without paying this exponential cost, and uses learned token embeddings so similar phenomena share statistical strength. On this specific dataset the additional expressive capacity of \phenseq{} does not produce a large advantage on aggregate metrics; it does produce a small advantage on exact gold-chain matches (13 vs 11) and on chain-level evidence depth.

More importantly, the ILM framework is \emph{not equivalent to \phenseq{} alone}: it is \phenseq{} (proposer) combined with \pheno{} (the corpus-grounded verifier with per-edge evidence, transition probabilities, and audit reports). The Markov ablation in this section replaces only the proposer; the verification layer where the deterministic-refusal and audit-trail properties live has no Markov analogue. A Markov chain has no notion of `is this edge supported by paragraph $X$ in paper $Y$', no notion of `refuse if no graph path exists', and no notion of a reproducible audit record. We retain the Markov ablation as the cleanest available test of this proposal/verification separation: a stripped-down proposer placed inside the same \pheno{}-grounded verification stack inherits the framework's auditability and refusal guarantees unchanged, which is the behaviour our contribution claim predicts. The proposer-level capabilities that distinguish \phenseq{} from a count-based table, full-prefix attention, endpoint conditioning, and embedding-based generalisation are the second locus of difference, and are characterised in the per-query analysis below.

For example, on the query \texttt{temperature\_gradient [GAP] mass\_transfer}, both proposers return the identical chain \texttt{temperature\_gradient} $\to$ \texttt{diffusion} $\to$ \texttt{mass\_transfer} because the graph-first router resolves the request before either proposer is invoked. The proposers diverge primarily on stress-suite queries where a multi-hop chain must be composed rather than recalled: Markov is more likely to terminate without reaching the goal, which inflates its correct-behaviour rate on adversarial inputs (since unreached goals count as correct rejections under our rubric) but offers no benefit on well-supported gold queries.

ILM diverges from any Markov-chain-based pipeline in three concrete ways, each of which is testable. (i)~Proposer capability: \phenseq{} attends over the full chain prefix and conditions on both endpoints in fill-in-the-middle mode (encoder input $(u, \textsc{gap}, v)$, decoder forced to start at $u$ and terminate at $v$; Sec.~\ref{sec:methods_phenoseq}), so it composes chains across endpoint pairs unseen during training. 86.1\% of gold-suite queries fall in this regime whereas a count-based Markov table has no mechanism for endpoint conditioning at all. (ii)~Learned similarity structure: \phenseq{}'s token embeddings let statistically similar phenomena share strength, so the proposer generalises across the heavy-tailed out-degree distribution of \pheno{}; a Markov table treats every phenomenon as an unrelated discrete state, and approximating long-range context with a higher-order $k$-gram would require $|V|^k$ parameters, infeasible for any $k \geq 3$. (iii)~Framework-level auditability: \pheno{} supplies the per-step evidence, deterministic refusal, and reproducible audit reports that define the contribution of ILM, and these properties have no analogue in a Markov-chain pipeline regardless of how the proposer is trained. The Markov ablation in this section tests only the proposer, deliberately, to establish that the auditability and refusal properties of the framework do not depend on proposer expressivity, they are supplied by \pheno{}, which is the principal claim of the paper. ILM is therefore not a benchmark improvement over a Markov chain but a framework with capabilities and guarantees that a Markov chain cannot, by construction, provide.

We interpret these results as supporting the framing of the ILM framework's contribution as the proposal/verification \emph{separation} rather than the seq2seq architecture per se. Any proposer that operates over the canonical phenomenon vocabulary and can be filtered against the graph admissibility mask is, in principle, a substitute for \phenseq{}. We retain \phenseq{} as the primary proposer because it provides the slight quality advantage on chain-level evidence depth and exact gold-chain matches that the data show, but a graph neural network or constrained search algorithm would be a viable alternative; the architectural commitment is to the verification-friendly proposer interface, not to encoder--decoder seq2seq.

\subsection{Comparison to vanilla LLM zero-shot}\label{sec:results_external}

To position the ILM pipeline against an external baseline, we evaluate vanilla Claude Opus 4.7 prompted zero-shot on the same 172 queries with chain-of-thought and ``cite supporting evidence'' instructions (full prompt in SI). The results reframe the contribution of the framework.

On gold queries, vanilla Claude Opus 4.7 achieves 91.7\% (33/36) pass rate, comparable to our pipeline's 94.4\% (34/36); the difference is not significant under Holm--Bonferroni correction. On out-of-vocabulary queries, vanilla Claude Opus 4.7 achieves 84.6\% correct-behaviour, exceeding our pipeline's 43.6\% by 41 points: the LLM trivially resolves \texttt{apparent\_viscosity} as a synonym for viscosity, while our alias-resolution layer is more conservative.

Two findings reveal structural distinctions that the comparable-pass-rate result obscures.

Figure~\ref{fig:arch_audit}(b) shows a concrete illustration. Given the deliberately impossible query \texttt{surface\_reaction} $\to$ \texttt{adsorption}, vanilla Claude Opus 4.7 produces a fluent 4-step chain (\texttt{surface\_reaction} $\to$ \texttt{reactant\_depletion\_at\_surface} $\to$ \texttt{concentration\_gradient\_formation} $\to$ \texttt{mass\_transport\_to\_surface} $\to$ \texttt{adsorption}) that reverses the canonical Langmuir--Hinshelwood causal direction; zero of the four consecutive transitions correspond to edges in \pheno{}. The ILM pipeline refuses the same query with \textsc{goal\_not\_reached} at Tier 1 because no graph path exists from \texttt{surface\_reaction} to \texttt{adsorption}.

First, the vanilla model fabricates a chain for every one of 31 deliberately-constructed missing-edge queries (0\% correct refusal). It never refuses. Our pipeline correctly refuses 26 of 31 (83.9\%) under the Tier-0/1/2 baseline and 29 of 31 (93.5\%) with the Claude Opus 4.7 referee. The same vanilla model, used as our Tier-3 referee, correctly identifies 3 of 5 implausible chains in this suite that the seq2seq stack accepts. We attribute the gap to the architectural difference: in our pipeline, ``no graph path exists'' is a structural input to the decision; in vanilla LLM inference, no such signal is available, and the model defaults to chain production whenever asked.

Second, the chains produced by vanilla Claude Opus 4.7 do not correspond to walks on our evidence graph. Of the consecutive $(u,v)$ transitions in vanilla LLM-produced chains across all 172 queries, only 0.70\% match edges in \pheno{}; 142 of 146 PASS chains contain zero graph-supported bigrams. The chains read as fluent physics narrative; citing real laws, real textbooks, and real papers in 81\% of an inspected sample (15 chains, 58 citation steps; full audit in the released citation-check artifact), but the chain steps themselves use phenomenon names that do not map to a controlled vocabulary, so the chain cannot be verified against any specific corpus. Every accepted chain in our pipeline, by construction, has 100\% graph-supported transitions traceable to specific corpus paragraphs.

We frame these two distinctions: deterministic refusal of evidence-absent queries, and graph-supported per-step traceability as the operational difference between corpus-grounded ILM reasoning and standard LLM generation. The framework does not claim pass-rate superiority on standard queries: vanilla LLMs are competitive on well-represented domains. The framework does claim two structural properties that vanilla LLMs lack and that, in our experiments, cannot be recovered through better prompting alone.

\subsection{Failure-mode composition and repair behaviour}\label{sec:results_failures}

Decomposing failed queries by primary cause revealed that the dominant residual failure mode was \textsc{goal\_not\_reached}, with the count remaining at 35 across all referee-enabled stress configurations. This confirms that the bottleneck is chain completion in weakly connected graph regions rather than per-edge scoring. The referee introduced an additional rejection mode (\textsc{referee\_implausible}), most prominent for Claude Opus 4.7, which rejected 7 gold-suite chains on plausibility grounds despite graph support.

For example, the query \texttt{higher\_pressure [GAP] microstructure} produces a graph-supported chain (\texttt{higher\_pressure} $\to$ \texttt{compression} $\to$ \texttt{microstructure}) but only narrowly: both edges carry support 1 (single corpus paragraph each), and the second edge ranks 6/7 among the alternatives leaving \texttt{compression}. A more densely-supported alternative routing through \texttt{porosity} (rank 1/7 from \texttt{compression}, support 3) is unreachable because no edge exists from \texttt{porosity} to \texttt{microstructure} in the current graph. This is the chain-completion bottleneck in weakly-connected graph regions: limited corpus coverage at one node forces the proposer onto lower-ranked successors.

Waypoints were offered broadly by the referee (80--88\% of stress queries) yet resolved fewer than 19\% of the queries that received them, as suggested intermediates often targeted phenomena that were themselves weakly connected in \pheno{}. The improvement from referee-on-fail arose from more effective search over candidate mediators rather than loosened evidence thresholds; repaired chains still had to survive graph-backed acceptance (Supplementary Fig.~S1).

\subsection{Evidence quality of accepted chains}\label{sec:results_evidence}

\begin{figure}[htbp]
    \centering
    \includegraphics[width=\linewidth]{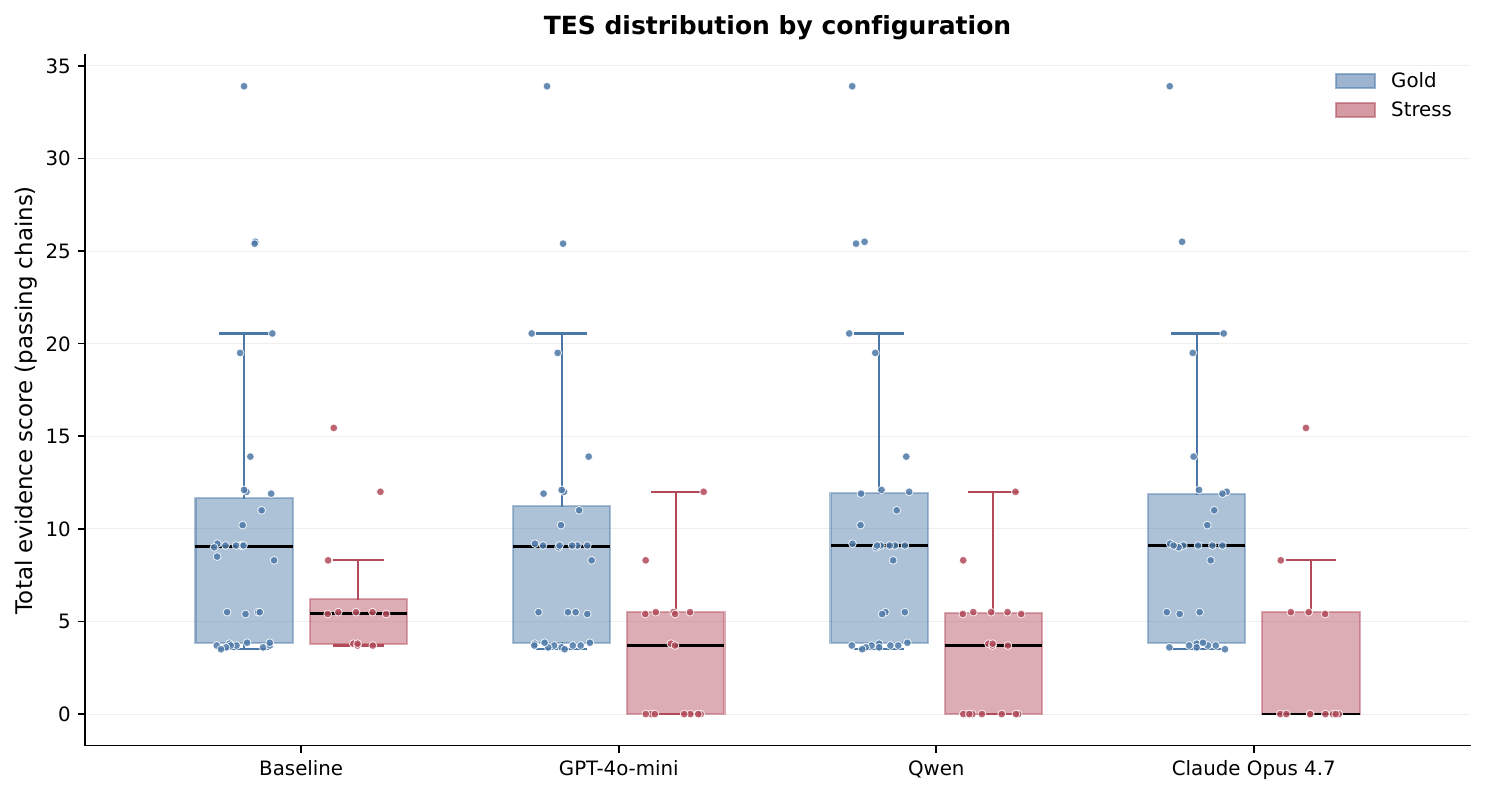}
    \caption {Distribution of Total Evidence Score across configurations. Individual accepted chains are shown as jittered points overlaid on box plots. Gold-suite chains (left) maintain stable evidence depth; stress-suite chains (right) carry lower but non-trivial evidence.}
    \label{fig:evidence_strip}
\end{figure}

Accepted chains were predominantly short (two-edge paths) but did not collapse to trivial or unsupported explanations. Gold-suite chains maintained stable evidence depth across configurations (mean TES $9.4$--$10.2$; per-edge support 2.28--2.50; per-edge weight 4.37--4.81). Stress-suite accepted chains carried lower but non-trivial evidence (mean TES 5.7--6.5), confirming that improvements did not arise from accepting unsupported explanations (Fig.~\ref{fig:evidence_strip}).

For example, the highest-evidence accepted chain in the gold suite is \texttt{temperature\_gradient} $\to$ \texttt{diffusion} $\to$ \texttt{mass\_transfer} (TES = 33.9): the first edge is supported by 12 corpus paragraphs (transition probability $p = 0.432$, ranked first of 15 outgoing alternatives) and the second by 5 paragraphs ($p = 0.116$, ranked first of 40). At the other extreme, low-evidence accepted chains such as \texttt{higher\_pressure} $\to$ \texttt{compression} $\to$ \texttt{microstructure} (TES = 3.6) rely on a single supporting paragraph per edge; these are accepted under the Tier-1 threshold but flagged in the per-step audit report (Sec.~\ref{sec:methods_audit}) for human review because both rank and support fall in the bottom quartile of the graph.

Figure~\ref{fig:evidence_strip} visualises the per-configuration TES distributions: gold-suite chains (left panel) cluster tightly around their stable mean across all four configurations, with the rightmost outlier at TES = 33.9 corresponding to the \texttt{temperature\_gradient} $\to$ \texttt{diffusion} $\to$ \texttt{mass\_transfer} chain. Stress-suite chains (right panel) show lower medians but a non-zero floor at per-edge support $\geq 2$, demonstrating that referee-on-fail repairs draw on evidence-backed alternatives rather than relaxing the evidence threshold. The visual stability of the gold-suite distributions across configurations supports the framework's claim that the verification layer maintains a hard evidence floor irrespective of proposer or referee choice.

\subsection{Local auditability of accepted chains}\label{sec:results_casestudy}

\begin{figure}[htbp]
    \centering
    \includegraphics[width=0.85\linewidth]{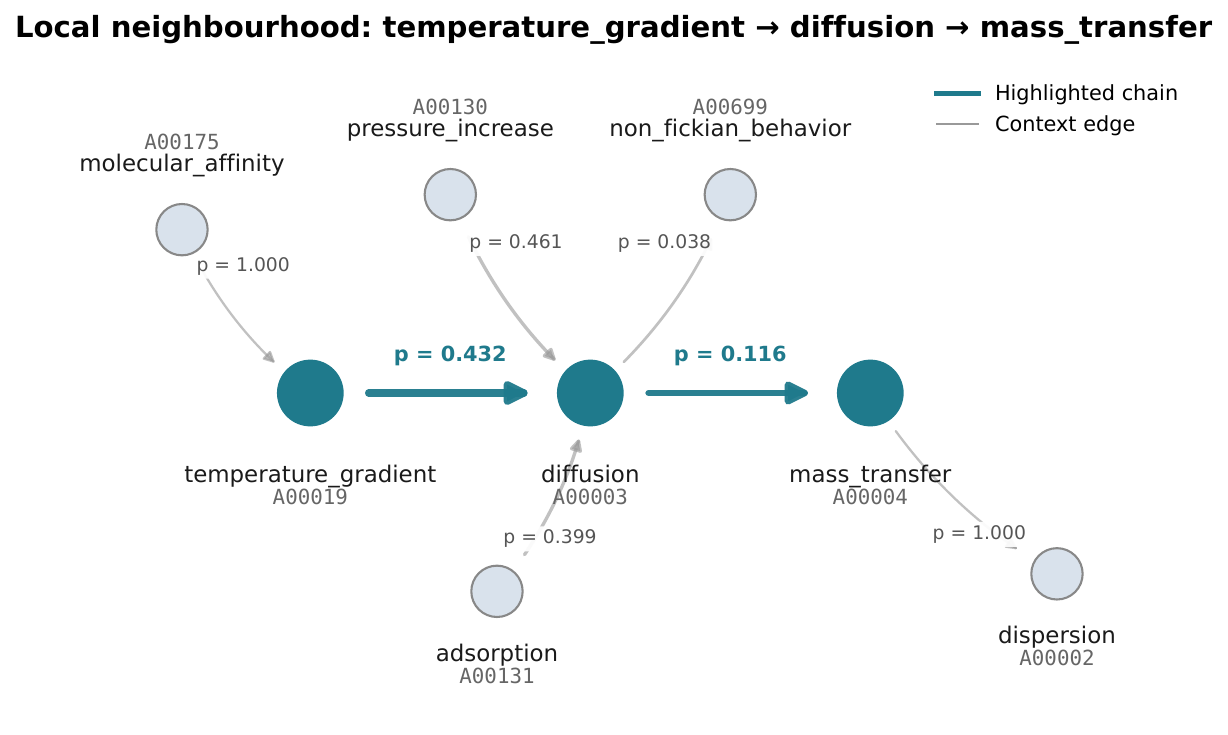}
    \caption{Local subgraph centred on an accepted endpoint chain. Edge widths are proportional to aggregated weight $w(u,v)$; transition probabilities are annotated. The highlighted path shows an auditable mechanism within the broader \pheno{} neighbourhood.}
    \label{fig:case_study}
\end{figure}

To illustrate what an audited explanation looks like in practice, Figure~\ref{fig:case_study} shows the local subgraph centred on the highest-evidence accepted chain in the gold suite, \texttt{temperature\_gradient} $\to$ \texttt{diffusion} $\to$ \texttt{mass\_transfer} (TES = 33.9; selected because it has the maximum aggregated evidence weight of any accepted gold chain). Edge widths are proportional to aggregated edge weight $w(u,v)$; transition probabilities $p(u \to v)$ are annotated on each edge. The highlighted path is embedded in a broader neighbourhood of nearby phenomena (\texttt{adsorption}, \texttt{pressure\_increase}, \texttt{non\_fickian\_behavior}, \texttt{dispersion}, \texttt{molecular\_affinity}) that are not part of the accepted chain but are reachable in one hop, illustrating that the returned explanation is a supported traversal through a reusable statistical knowledge object rather than an isolated string. The same audit format, per-step transition probability, support count, ranked alternatives, and verbatim supporting paragraphs applies to every accepted chain in the pipeline, providing a uniform inspection surface for downstream review. This worked example makes the auditability claim tangible: a single accepted chain comes with a structured neighbourhood of alternatives, edge-weight magnitudes, and explicit paragraph-level evidence anchors that a reviewer can inspect end-to-end without re-running the pipeline.
Figure~\ref{fig:case_study} shows a representative local subgraph centred on an accepted endpoint chain, with edge widths proportional to aggregated weight. The highlighted chain is embedded in a broader neighbourhood of nearby phenomena, demonstrating that returned explanations are supported traversals through a reusable statistical knowledge object, not isolated strings. Supplementary Figs.~S2--S4 presents qualitative examples spanning gold-suite successes (including the highest-evidence chain: TES\,=\,33.9, 12 supporting paragraphs, referee score 0.93) and correct stress-suite behaviour including rejection of physically implausible queries.

\subsection{Concrete examples: same query across all configurations}\label{sec:results_concrete}

To illustrate how the pipeline behaves in practice, we present verbatim system outputs for three representative queries (two from the gold suite and one from the stress suite) evaluated under all four configurations. The full outputs are reproduced in Supplementary Figs.~S2--S4.

 \paragraph{Demonstrating the fill-in capability.}
A central function of the ILM framework is its ability to \emph{fill in missing intermediate phenomena} between two endpoints under specific physical conditions. This goes beyond simple retrieval: the system must identify which phenomena plausibly mediate between cause and effect, even when no direct edge connects them. For the query \texttt{higher\_pressure [GAP] diffusion}, no direct edge exists from higher\_pressure to diffusion in the graph. The pipeline must infer that higher pressure leads to compression of the porous medium, which alters its porosity, which in turn modifies the effective diffusion rate. The resulting three-step chain: higher\_pressure $\rightarrow$ compression $\rightarrow$ porosity $\rightarrow$ diffusion, is constructed by \phenseq{} proposing candidate intermediates and \pheno{} verifying that each step has corpus support (support counts of 1, 3, and 2 respectively; TES\,=\,11.0). This fill-in capability distinguishes the framework from both keyword retrieval (which cannot infer intermediates) and standard LLMs (which can propose intermediates but cannot verify them against evidence). Supplementary Fig.~S3 shows the verbatim outputs across all configurations for this query.

Given the query \texttt{temperature\_gradient [GAP] mass\_transfer}, all four configurations produce the identical two-step chain: temperature\_gradient $\rightarrow$ diffusion $\rightarrow$ mass\_transfer (Supplementary Fig.~S2). The chain achieves TES\,=\,33.9, the highest in the entire gold suite, backed by 12 supporting corpus paragraphs for the first step (temperature\_gradient $\rightarrow$ diffusion, $p = 0.432$, weight = 23.6) and 5 for the second (diffusion $\rightarrow$ mass\_transfer, $p = 0.116$, weight = 10.3). The baseline returns this chain through graph-first routing without invoking any generation model. The three referee configurations accept the same chain but provide different plausibility scores of 0.80 (GPT-4o-mini), 0.85 (Qwen), and 0.93 (Claude Opus 4.7), with qualitatively different rationales. Claude Opus 4.7 provides the most mechanistically detailed justification, noting that ``temperature gradients change molecular kinetic energy differentials and transport coefficients'' and suggesting diffusion\_coefficient and concentration\_gradient as additional waypoints. This example demonstrates that the graph-backed core is sufficient for well-supported queries, while the referee adds interpretive depth without changing the accepted chain.

For the query \texttt{pressure\_increase [GAP] mass\_transfer}, all configurations again produce the same chain: pressure\_increase $\rightarrow$ diffusion $\rightarrow$ mass\_transfer (TES\,=\,20.6; Supplementary Fig.~S3). Here, referee scores diverge more substantially: GPT-4o-mini and Qwen both score 0.90, while Claude Opus 4.7 scores only 0.67, noting that ``a pressure increase can contribute to diffusion indirectly by creating pressure or concentration gradients, but pressure alone is not usually the direct generic cause of diffusion.'' This illustrates how the referee can flag \emph{physically nuanced} concerns even when the graph evidence is strong, providing a complementary signal for expert review.

For the stress query \texttt{clustering [GAP] mass\_transfer}, the correct behaviour is rejection, because clustering does not physically lead to mass transfer. The corpus contains no evidence supporting this causal direction (Supplementary Fig.~S4). All four configurations correctly reject this query. The baseline fails at Tier~1 (GOAL\_NOT\_REACHED) because no evidence-supported path from clustering to mass\_transfer exists. The three referee configurations independently confirm the rejection with low plausibility scores of 0.10 (GPT-4o-mini), 0.12 (Claude Opus 4.7), and 0.20 (Qwen), and provide rationales explaining the physical implausibility. This example shows the system's ability to enforce physical directionality and refuse to hallucinate unsupported mechanistic pathways.

\subsection{Tier-3 referee ablation}\label{sec:results_referee}

Despite limited repair effectiveness, all three referee models assigned substantially higher plausibility scores to chains that ultimately passed verification than to those that failed ($\Delta = 0.34$--$0.47$), indicating well-calibrated quality assessment (Table~\ref{tab:discrim}). GPT-4o-mini provided the strongest overall balance between gold-suite stability and stress-suite robustness. The Qwen referee produced comparable behaviour in the on-fail configuration, suggesting that the repair mechanism itself plays a larger role than the specific model architecture. The Claude Opus 4.7 referee's aggressive plausibility filtering degraded gold-suite performance by rejecting 7 valid graph-supported chains.

The Tier-3 referee variants exhibit a consistent calibration trade-off relative to the Tier-0/1/2 baseline. On the gold suite, every referee variant reduces pass rate (Claude Opus 4.7: 80.6\%, GPT-4o-mini: 88.9\%, Qwen: 86.1\%) relative to the 94.4\% baseline, because each referee occasionally rejects valid graph-supported chains as physically implausible. On the stress suite (Table~\ref{tab:stress_correct}), the same referees reduce overall correct-behaviour from 69.7\% (baseline) to 59.1--68.2\%, because the referees override the structural self-loop FAIL signal. Within referee variants, however, Claude Opus 4.7 substantially improves correct-behaviour on specific categories: dead-end queries rise from 60\% (baseline) to 100\%, and missing-edge queries (Sec.~\ref{sec:results_oov_missing}) rise from 83.9\% to 93.5\%. We therefore retain the referee as an optional ablation that adds category-specific value rather than universal improvement, consistent with the framing of Tier-3 as outside the auditable core.

\begin{table}[htbp]
\centering
\caption{Referee plausibility scores stratified by chain outcome (stress suite). All three models assign substantially higher scores to passing chains, indicating well-calibrated quality assessment.}
\label{tab:discrim}
\small
\setlength{\tabcolsep}{5pt}
\begin{tabular}{@{}l cc c@{}}
\toprule
& \multicolumn{2}{c}{\textbf{Mean plausibility score $\pm$ s.d.}} & \\
\cmidrule(lr){2-3}
\textbf{Referee model}
 & PASS chains
 & FAIL chains
  & $\Delta$ \\
\midrule
GPT-4o-mini
  & $0.83 \pm 0.08$ \;($n{=}9$)
  & $0.42 \pm 0.23$ \;($n{=}48$)
 & 0.41 \\[3pt]
Qwen
  & $0.72 \pm 0.08$ \;($n{=}11$)
  & $0.25 \pm 0.09$ \;($n{=}47$)
 & \best{0.47} \\[3pt]
Claude Opus 4.7
  & $0.71 \pm 0.10$ \;($n{=}13$)
  & $0.27 \pm 0.13$ \;($n{=}41$)
 & 0.44 \\
\bottomrule
\end{tabular}
\end{table}

\section{Discussion}\label{sec:discussion}

The contribution of the ILM framework is the architectural separation between statistical chain proposal (\phenseq{}, over a closed vocabulary of physical phenomena) and corpus-grounded chain verification (\pheno{}, with per-step evidence trail). This separation makes the verification layer reusable as a structured artifact independent of the proposal model: any candidate chain, whether generated by \phenseq{}, by an external LLM, by a 1st-order Markov chain (Sec.~\ref{sec:results_ablation}), or by a domain scientist, can be checked against the same graph and the same audit format. Compared with retrieval-augmented systems including GraphRAG~\cite{c1799bf28d1ae93e1631be5b59196ee1e568f538} and CausalRAG~\cite{71bb53ceb14affbac0811516dfa6d22f49f7dda6}, which retrieve passages at inference time to condition a generative model, \pheno{} pre-compiles a structured statistical object that serves as a reusable verification layer. Because this object is fixed once compiled, the per-step audit values returned for a given query are deterministic and reproducible across runs.

This separation has concrete consequences for auditability and reliability. When a standard LLM states a connection between two concepts, its output does not inherently include a deterministic audit mechanism showing whether the claim is supported by 1 or 100 corpus passages, how the transition scores relative to alternatives, or whether the reverse direction is equally supported. Tool-augmented or retrieval-augmented LLM pipelines can recover such grounding at inference time, but at the cost of run-to-run variability in retrieved evidence. \pheno{} addresses these limitations explicitly under Tiers~0--2. For the query \texttt{temperature\_gradient [GAP] mass\_transfer}, the pipeline does not merely state a connection; it returns a chain (temperature\_gradient $\rightarrow$ diffusion $\rightarrow$ mass\_transfer) with $p = 0.432$ for the first step (ranked 1st among 15 outgoing edges, backed by 12 corpus paragraphs), $p = 0.116$ for the second step (ranked 1st among 40 outgoing edges, backed by 5 paragraphs), and a chain-level TES of 33.9. Every number is reproducible because it is computed from a fixed graph artifact derived from corpus statistics. The pipeline's ability to refuse unsupported queries (Sec.~\ref{sec:results_oov_missing}: 93.5\% correct refusal of missing-edge queries with the Tier-3 referee, vs.\ 0\% for vanilla Claude Opus 4.7 zero-shot on the same queries) is correct by graph construction: a chain to a goal that has no supported path in the graph cannot be produced. The handling of self-loops is similarly structural. We retain these results as a confirmation that the graph-structural checks are correctly integrated, not as evidence of learned causal generalisation.

The fundamental distinction between an Interaction Language Model and a standard Large Language Model lies not in a broad claim that LLMs cannot reason, but in what is being modeled, constrained, and verified statistically. A standard LLM learns statistical relationships between words or subword tokens: it captures how tokens co-occur in text corpora and generates fluent continuations based on these language-level patterns. When an LLM produces the sentence “a temperature gradient increases diffusion rate,” the output may be scientifically plausible, but the model itself does not necessarily expose a fixed evidence path showing where that physical relationship is supported, nor is it forced to refuse the claim when such evidence is absent. The ILM framework, by contrast, learns statistical relationships between physical phenomena themselves. Each node in the interaction graph represents a canonical phenomenon rather than a word, and each edge encodes how frequently and confidently one phenomenon leads to another in the scientific literature. The resulting statistics are interaction statistics: they describe how physical processes depend on one another in real systems, grounded in paragraph-level corpus evidence rather than only in token co-occurrence. This corpus-derived verification layer gives scientific chain reasoning two properties that ordinary prompting does not guarantee: deterministic refusal when no admissible evidence supports a proposed step, and per-step auditability when evidence is present. Thus, the contribution of the ILM is not that it universally outperforms frontier LLMs, GraphRAG, KG-RAG, or tool-augmented systems, but that it provides a fixed, evidence-grounded mechanism for constructing, verifying, and rejecting chains of physical phenomena.

This difference has concrete consequences for auditability and reliability. When a standard LLM states a connection between two concepts, there is no mechanism to verify whether this claim is supported by 1 or 100 corpus passages, what the transition probability is relative to alternatives, or whether the reverse direction is equally supported. \pheno{} addresses each of these limitations explicitly. For the query \texttt{temperature\_gradient [GAP] mass\_transfer}, the pipeline does not merely state a connection; it returns a chain (temperature\_gradient $\rightarrow$ diffusion $\rightarrow$ mass\_transfer) with $p = 0.432$ for the first step (ranked 1st among 15 outgoing edges, backed by 12 corpus paragraphs), $p = 0.116$ for the second step (ranked 1st among 40 outgoing edges, backed by 5 paragraphs), and a chain-level TES of 33.9. Every number is reproducible because it is computed from a fixed graph artifact derived from interaction statistics, not sampled from a stochastic word-level model. Furthermore, the pipeline's ability to \emph{reject} unsupported queries, as demonstrated by the 100\% rejection rate for reversed-edge and self-loop queries, has no analogue in standard LLM inference, where the model will always produce \emph{some} output regardless of whether the requested reasoning direction is physically valid. This rejection capability arises directly from the interaction-statistical foundation: because the graph encodes directed physical dependencies rather than symmetric word associations, it can distinguish ``A causes B'' from ``B causes A'' in a way that word-level models cannot.

The results demonstrate that separating proposal and verification layers enables robust and interpretable mechanistic reasoning. The Tier~0/1/2 baseline achieves 94.4\% accuracy on in-distribution queries without any LLM referee, confirming that the graph-backed core pipeline is sufficient when the endpoint relation is well represented in the interaction-statistical knowledge layer. The apparently low stress-suite pass rates (18--29\%) reflect the system correctly refusing to produce chains where no valid mechanistic path exists, as confirmed by 70--80\% correct-behaviour rates. This behaviour contrasts sharply with word-level LLMs, which cannot distinguish ``no valid path exists'' from ``I lack the vocabulary to express it.''.

The complementary roles of \pheno{} and \phenseq{} are central to the framework, and both operate on interaction statistics rather than word statistics. \phenseq{} proposes candidate chains over phenomenon tokens whose transition probabilities reflect physical co-occurrence in the literature; \pheno{} verifies those chains through explicit graph structure, transition probabilities, and supporting evidence. The referee-on-fail configurations demonstrate that targeted mediator suggestions can improve recovery on difficult queries without relaxing evidence constraints; the referee acts as a navigation aid, and repaired chains must still satisfy the evidential requirements imposed by \pheno{}.

 The strong score separation between PASS and FAIL chains ($\Delta = 0.34$--$0.47$) suggests that referee scores could serve as a useful reranking signal in future iterations, even when waypoint suggestions do not directly resolve queries. However, the Claude Opus 4.7 referee's aggressive filtering degraded gold-suite performance (7 valid chains rejected), indicating that using referee scores as a soft signal rather than a hard filter could preserve discriminative capability while reducing false positives. The principal advantage of the referee lies in structured reasoning guidance rather than raw model capacity: once mediating phenomena are proposed, \pheno{} verification determines whether resulting chains satisfy evidential requirements.

 We note that \pheno{} is not a retrieval-augmented generation system: rather than retrieving passages at inference time to condition a generative model, we pre-compute a structured statistical object (the phenomenon interaction graph with edges, probabilities, and evidence) that serves as a reusable verification layer. Because this object encodes interaction statistics between physical phenomena rather than word-level retrieval scores, the audit trail is deterministic and reproducible (the same graph always produces the same per-step scores), whereas retrieval-augmented approaches may return different passages across runs, making verification less stable. Future work could explore hybrid architectures in which \pheno{} serves as the verification backend for retrieval-augmented or fine-tuned generative models.

The correct-behaviour analysis identifies a single clear limitation shared across all configurations: rare-token generalisation (0\% for all 8 queries), while every other adversarial category achieves 60--100\% correct behaviour. This precisely scoped failure mode points to graph densification (through expanded corpus coverage, cross-domain transfer, or active evidence acquisition) as the most impactful avenue for improvement. Additionally, the current study focuses on gas diffusion phenomena; further work will determine how well the framework generalises to broader physical reasoning domains. The OOV/Alias test suite ($n=39$) and Missing-Edge test suite ($n=31$) provide ready-made evaluation axes for these extensions.

Because \pheno{} records transition probabilities, supporting evidence, and chain-level summaries derived from interaction statistics in the physical domain, successful explanations can be traced to specific mechanistic relationships rather than being justified solely by model confidence or word-level plausibility. This structure enables failures to be localised to distinct causes (missing graph support, weak proposal chains, or plausibility rejection by the referee), providing a foundation for auditable scientific reasoning systems that goes beyond what either purely generative word-level models or traditional knowledge graphs currently offer.


\section{Conclusion}\label{sec:conclusion}

%

We introduced the Interaction Language Model (ILM) concept, treating physical interactions as a structured, learnable language; and demonstrated it through \pheno{} and \phenseq{} applied to molecular diffusion. \pheno{} converts domain literature into an auditable directed graph over phenomena with evidence-weighted edge scores; \phenseq{} proposes and ranks missing mechanistic steps under endpoint constraints, with proposals verified against the graph's evidence layer. The framework provides two structural properties that vanilla LLMs lack on the same evaluation queries: deterministic refusal of evidence-absent mechanisms and per-step audit traceability to specific corpus paragraphs. The framework is, in principle, applicable to wide range of scientific domains in which function emerges from sequential, conditional, evidence-grounded interactions among basic entities. Validating this generality requires constructing analogous interaction and phenomenon vocabularies as well as corpora in each domain and is left as future work.

\backmatter
\section*{Acknowledgements}

This work was supported by the \textbf{Gordon and Betty Moore Foundation}, grant DOI 10.37807/GBMF12246. The authors acknowledge the interdisciplinary research environment provided by the School of Computing and the Department of Physics and Astronomy at the University of Georgia, which supported the development of the Interaction Language Model framework presented in this study.

\subsection*{Data availability}

The data supporting the findings of this study are provided in the Supplementary Information.

\subsection*{Code availability}

The code supporting the findings of this study is provided in the Supplementary Information.

\subsection*{Generative AI and LLM use}
Large language models were used as research tools in the relation-extraction and referee components described in Methods, with model identifiers, prompts, caches, and audit logs reported for reproducibility.


\bibliography{referencesLargeInteractionmodel}

\end{document}